\documentclass[twocolumn,english,superscriptaddress,floatfix,longbibliography]{revtex4-1}
\usepackage[T1]{fontenc}
\usepackage[utf8]{inputenc}
\usepackage{xcolor}
\usepackage{pdfcolmk}
\usepackage{float}
\usepackage{amsmath}
\usepackage{amssymb}
\usepackage{graphicx}
\PassOptionsToPackage{normalem}{ulem}
\usepackage{ulem}

\makeatletter

\providecommand{\tabularnewline}{\\}
\providecolor{lyxadded}{rgb}{0,0,1}
\providecolor{lyxdeleted}{rgb}{1,0,0}

\DeclareRobustCommand{\lyxdeleted}[3]{{\color{lyxdeleted}\lyxsout{#3}}}
\DeclareRobustCommand{\lyxsout}[1]{\ifx\\#1\else\sout{#1}\fi}

\usepackage{placeins}

\makeatother

\usepackage{babel}
\begin{document}
\title{Critical growth of cerebral tissue in organoids: theory and experiments}
\author{Egor I. Kiselev}
\altaffiliation{E.K. and F.P. contributed equally to this work}

\affiliation{Center for Integrative Bioinformatics Vienna (CIBIV), Max Perutz Laboratories,
University of Vienna and Medical University of Vienna, Vienna Bio
Center (VBC), Vienna, Austria}
\affiliation{Physics Department, Technion, 320003 Haifa, Israel}
\author{Florian Pflug}
\altaffiliation{E.K. and F.P. contributed equally to this work}

\affiliation{Center for Integrative Bioinformatics Vienna (CIBIV), Max Perutz Laboratories,
University of Vienna and Medical University of Vienna, Vienna Bio
Center (VBC), Vienna, Austria}
\affiliation{Biological Complexity Unit, Okinawa Institute of Science and Technology
Graduate University, Onna, Okinawa 904-0495, Japan}
\author{Arndt von Haeseler}
\affiliation{Center for Integrative Bioinformatics Vienna (CIBIV), Max Perutz Laboratories,
University of Vienna and Medical University of Vienna, Vienna Bio
Center (VBC), Vienna, Austria}
\affiliation{Bioinformatics and Computational Biology, Faculty of Computer Science,
University of Vienna, Vienna, Austria}
\begin{abstract}
We develop a Fokker-Planck theory of tissue growth with three types
of cells (symmetrically dividing, asymmetrically dividing and non-dividing)
as main agents to study the growth dynamics of human cerebral organoids.
Fitting the theory to lineage tracing data obtained in next generation
sequencing experiments, we show that the growth of cerebral organoids
is a critical process. We derive analytical expressions describing
the time evolution of clonal lineage sizes and show how power-law
distributions arise in the limit of long times due to the vanishing
of a characteristic growth scale. We discuss that the independence
of critical growth on initial conditions could be biologically advantageous.
\end{abstract}
\maketitle

\paragraph*{Introduction}

The mechanisms of tissue growth and renewal are a core topic of stem-cell
research \citep{Chatzeli_2020_tracing_fate_stem_cells,Simons2011_cell_renewal_strategies}.
In particular, the role of stochasticity in cell differentiation is
discussed \citep{lin_2021_neural_progenitors_mammalian,zechner2020_stochasticity_differentiation,Klingler2020_cortical_stochastic,Llorca2019_stochastic_differentiation_cortex,Corominas2020_intestinal_niches,smith_2015_stochasticity_cell_diff}.
Single cell sequencing \citep{kashima_2020_single_cell_seq_rev},
combined with the labeling of cells with inheritable DNA sequences,
enables large scale, quantitative studies of cell populations in biological
tissues, where offspring populations can be traced back to their individual
ancestral cells \citep{wagner_2020_sc_omics_lineage_tracing,kester_2018_single_cell_lineage_tracing}.
Such lineage tracing experiments have revealed that offspring numbers
in mammalian cerebral tissue can vary by several orders of magnitude
\citep{Llorca2019_stochastic_differentiation_cortex,esk_2020_organoid_tissue_screen},
which supports the hypothesis that stochasticity is an important property
of cell proliferation and differentiation in the developing cerebral
cortex.

This work presents a study of lineage tracing data obtained by sequencing
15 cerebral organoids at different stages of their development \citep{esk_2020_organoid_tissue_screen}.
Cerebral organoids are highly controllable, self organized in vitro
models of the human cerebral cortex grown from stem cells. Organoids
are unique because they model human tissues which cannot be studied
in vivo. They have therefore become important biological tools to
study neural development and brain diseases \citep{lancaster2013_organoids,lancaster2017_organoids,kim2020human}.
We take a physics point of view on the population dynamics of cell
lineages in cerebral organoids, show that organoid growth is a critical
process, and discuss the biological implications.

\begin{figure}
\centering{}\includegraphics[width=0.4\columnwidth]{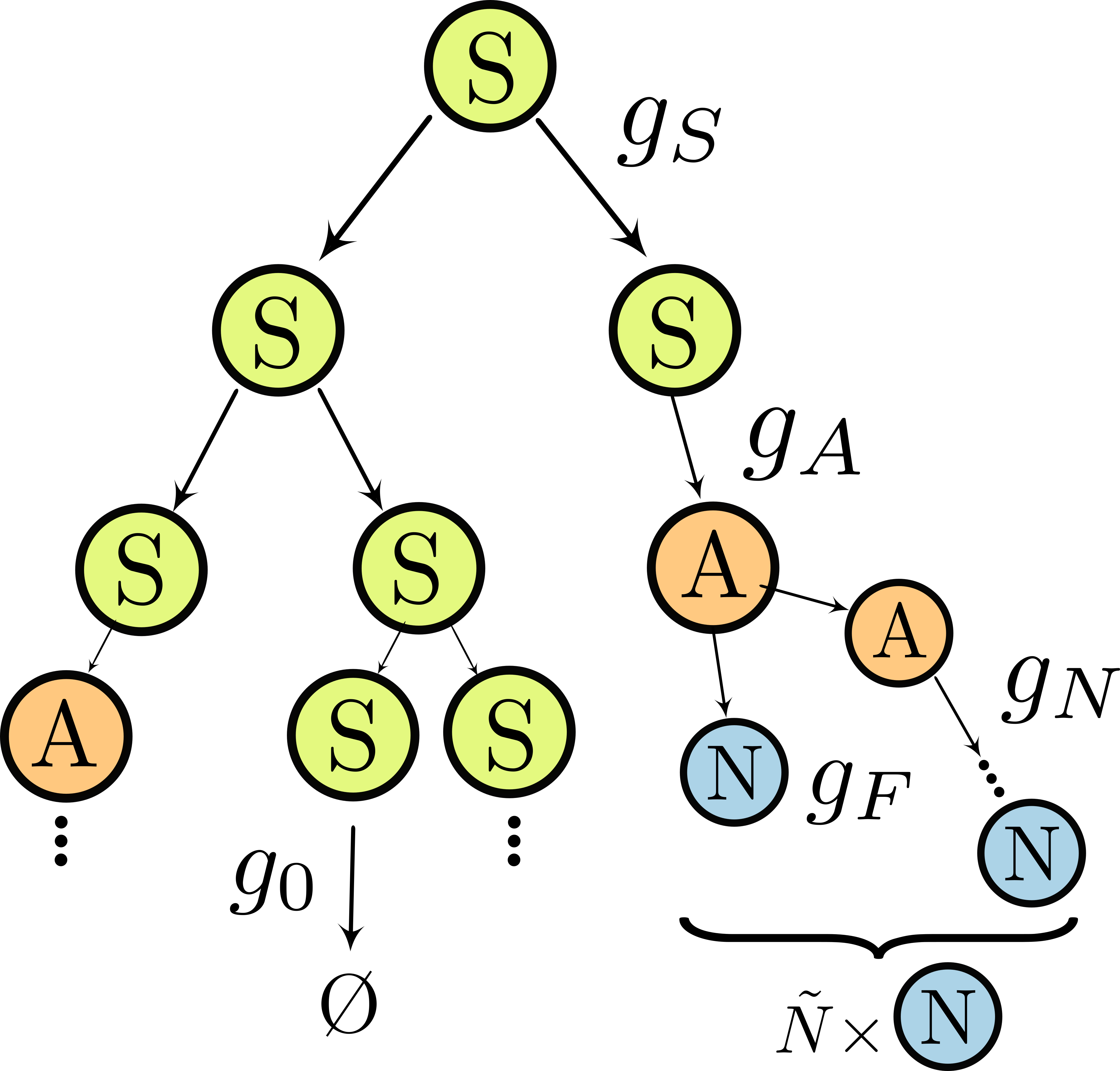}\caption{The SAN model of dividing and differentiating cells. Stem cells (green,
S) either divide at rate $g_{S}$ (S$\rightarrow$2S), differentiate
at rate $g_{A}$ (S$\rightarrow$A) or die at a rate $g_{0}$ (S$\rightarrow$$\text{Ø}$).
Differentiated cells (orange, A) that committed to a developmental
trajectory either divide asymmetrically (A$\rightarrow$A+N) at a
rate $g_{A}$, or differentiate directly into non-dividing N cells
(rate $g_{F}$). On average, each A cell produces $\bar{N}$ N-cells
until it looses the ability to divide. At criticality, the rates $g_{S}$
(division) and $g_{A}$ (differentiation) are equal.\label{fig:SAN_model}}
\end{figure}

\paragraph*{Model}

\begin{figure*}
\centering{}\includegraphics[width=0.34\textwidth]{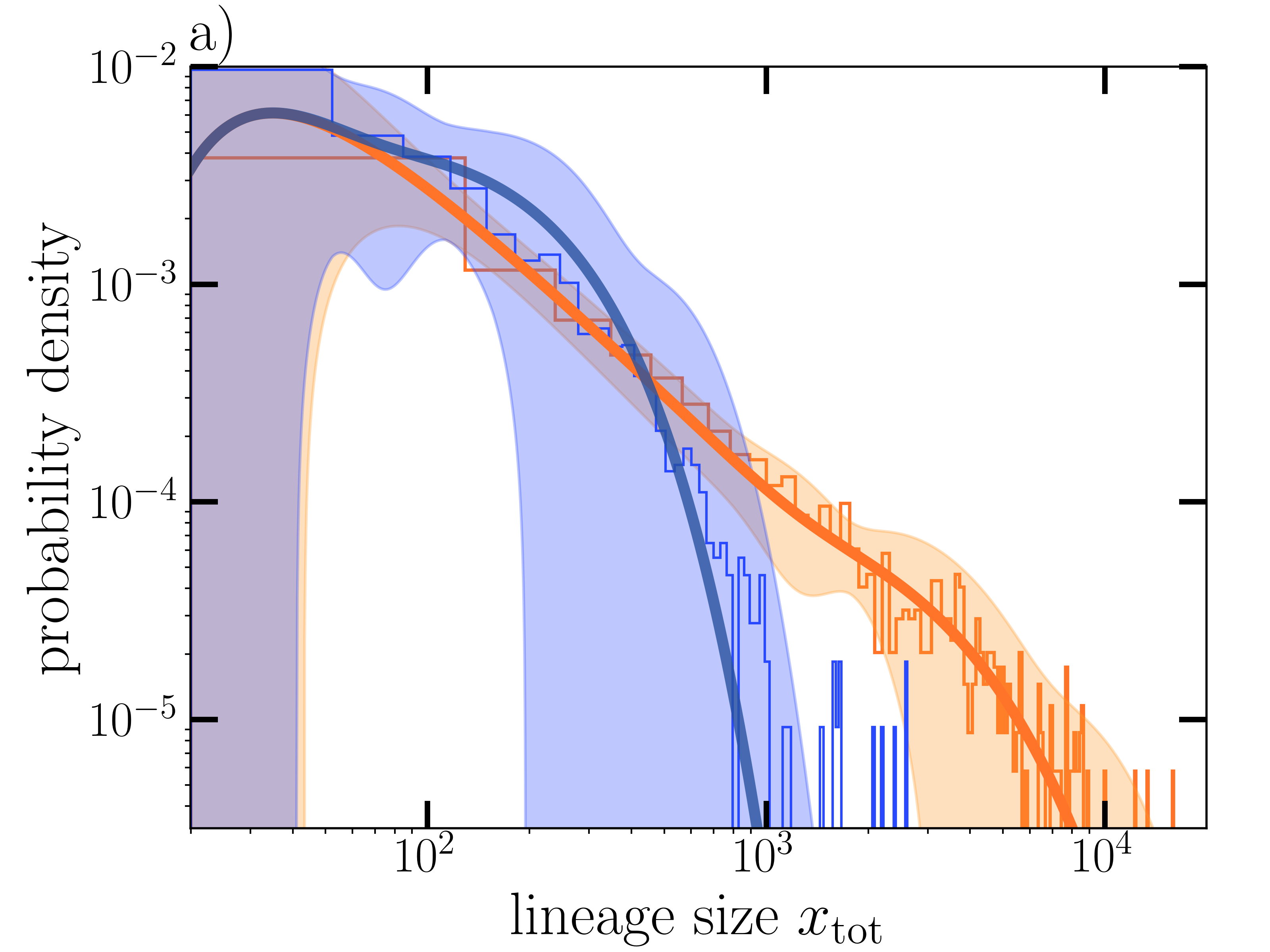}\includegraphics[width=0.34\textwidth]{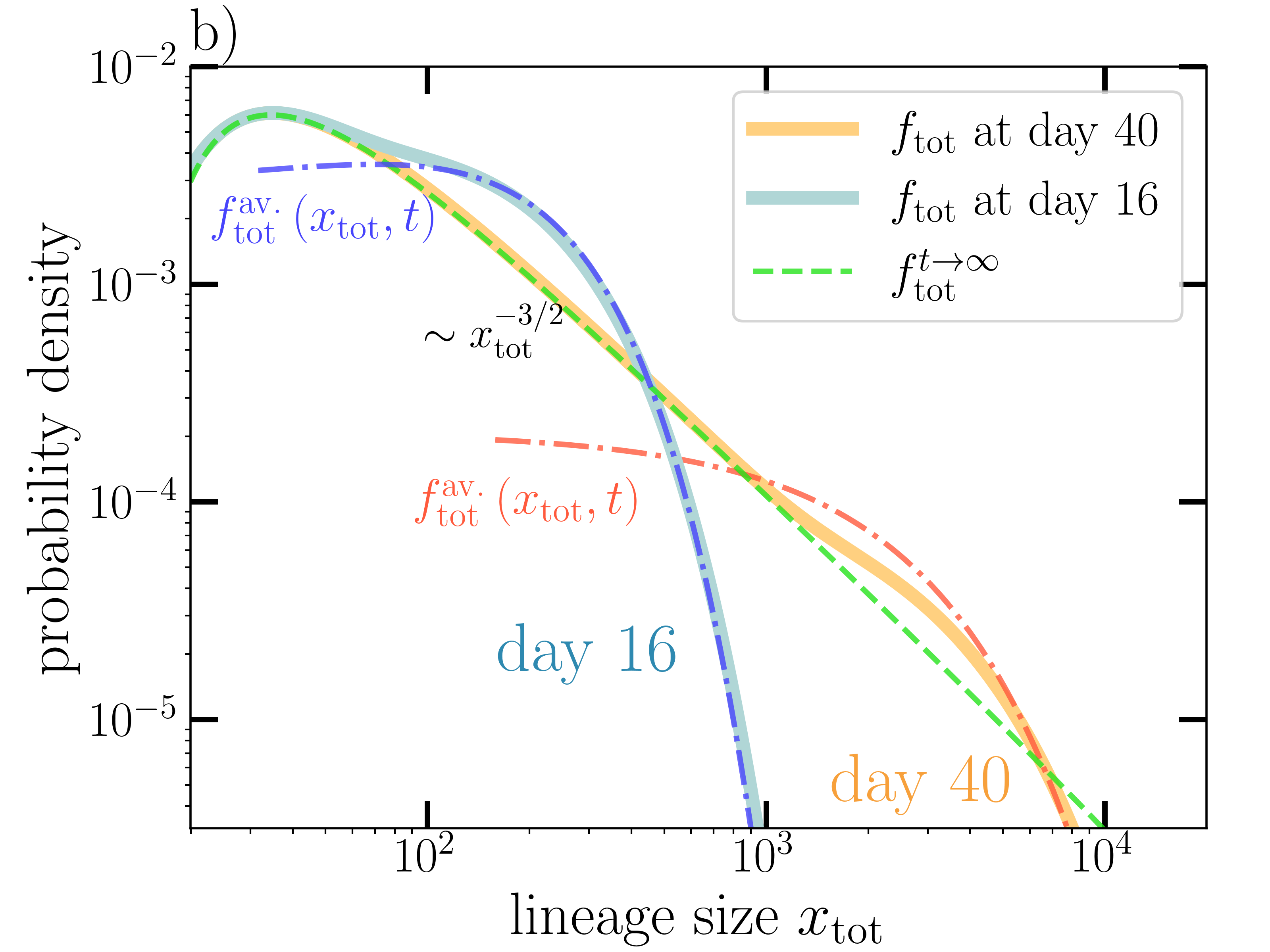}\includegraphics[width=0.34\textwidth]{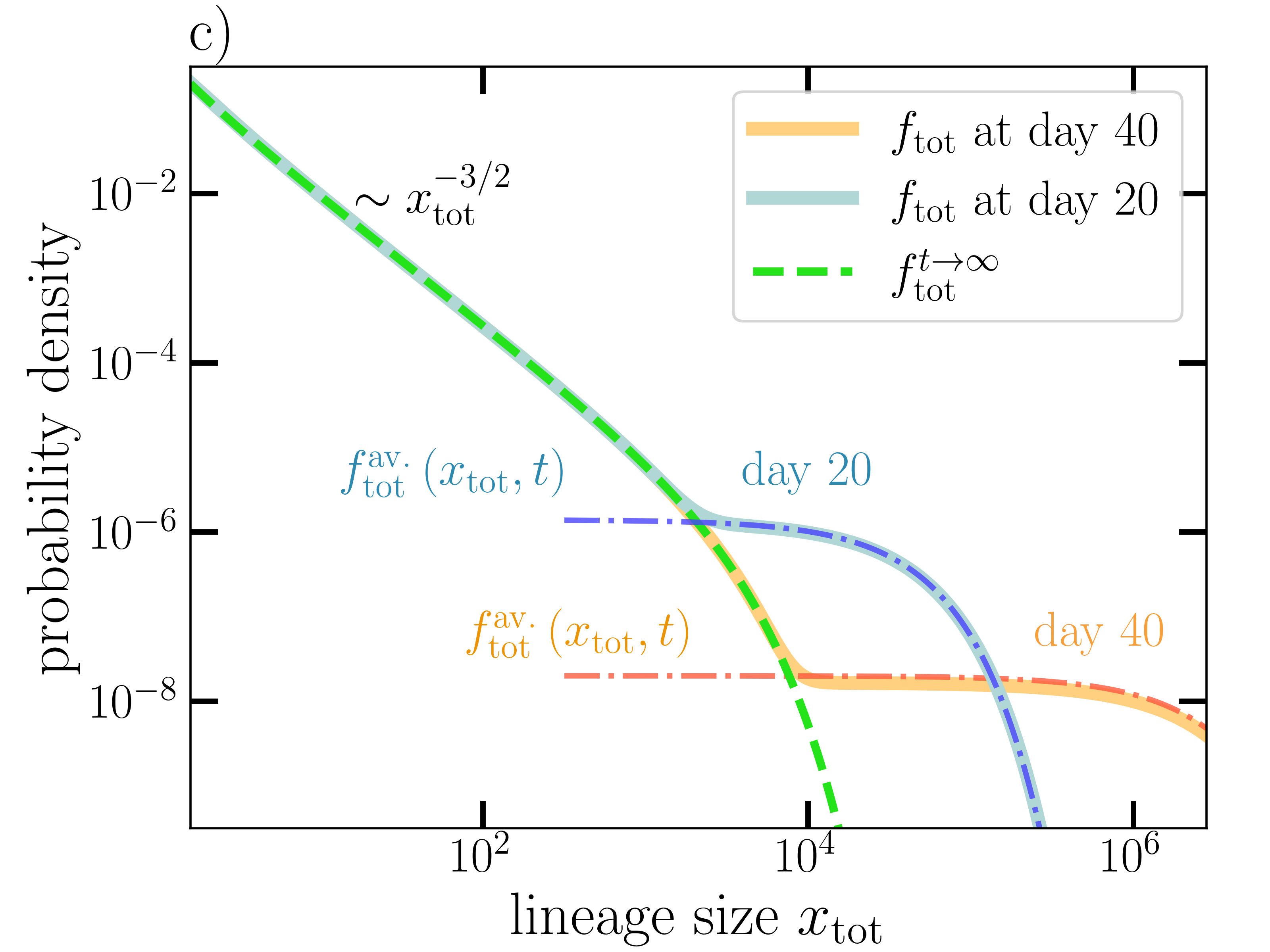}\caption{a) Histograms of lineage sizes of two organoids sequenced at $t=16\,\mathrm{days}$
(blue) and $t=40\,\mathrm{days}$ (orange) and the corresponding probability
densities of the SAN model $f_{\mathrm{tot}}\left(x_{\mathrm{tot}},t\right)$
(thick solid lines). Parameter values are given in Table \ref{tab:parameter_table}.
Shaded areas indicate error margnis. b) Theoretical SAN model probability
densities and the analytical approximations $f_{\mathrm{tot}}^{t\rightarrow\infty}\left(x_{\mathrm{tot}}\right)$
(dashed, green line) and $f_{\mathrm{tot}}^{\mathrm{av.}}\left(x_{\mathrm{tot}},t\right)$
(dashed, dotted lines) of Eqs. (\ref{eq:critical_Levy_distr}), (\ref{eq:avalanche_approx})
for the parameter estimates of Table \ref{tab:parameter_table} ($\alpha<0$
- sub-critical regime). For $t\rightarrow\infty$ the distribution
approaches the weakly truncated $3/2$-power-law Lévy distribution
$f_{\mathrm{tot}}^{t\rightarrow\infty}\left(x_{\mathrm{tot}}\right)$
everywhere. c) SAN model predictions for small $\alpha>0$ (super-critical
regime). $f_{\mathrm{tot}}\left(x_{\mathrm{tot}},t\right)$ still
approaches $f_{\mathrm{tot}}^{t\rightarrow\infty}\left(x_{\mathrm{tot}}\right)$,
except for very large lineage sizes, where the avalanche of active
S-cell proliferation dominates. We used $\alpha=0.2$, $\beta=10$,
$s_{0}=1$ and $N=1$. \label{fig:theory_data_powerlaws_avalanches}}
\end{figure*}

Our study begins with the observation that the numbers of descendants
of an individual stem cell in the organoid (lineage sizes) are roughly
distributed according to a 3/2-power-law. This behavior becomes more
and more pronounced at late stages of the organoid development. To
model the growth process mathematically, we build upon the theory
of continuous state branching processes \citep{feller1951_Two_singular_diffusion_problems,lamperti1967limit},
pioneered by Feller \citep{Feller_Kampf_ums_Dasein}. These processes
are known to lead to power-law distributed poluation sizes \citep{Zapperi1995_SOC,alava2012branching,diSanto_munoz_2017branching}. 

We thus introduce the \textit{SAN model}. It consists of three agents:
symmetrically dividing S-cells that represent stem cells, asymmetrically
dividing A-cells and non-dividing N-cells (fully developed cells,
e.g. neurons). S-cells undergo symmetric division (S$\rightarrow$2S)
at rate $g_{S}$, differentiation (S$\rightarrow$A) at rate $g_{A}$
and death (S$\rightarrow$0) at rate $g_{0}$. A-cells have committed
to a developmental trajectory and produce N-cells through asymmetric
divisions (A$\rightarrow$A+N) at a rate $g_{N}$ until the process
is terminated by direct differentiation (A$\rightarrow$N) (rate $g_{F}$).
The branching process of symmetric division and differentiation of
S-cells with rates $g_{S}$ and $g_{A}$ is at the heart of the SAN-model.
Criticality is reached, when the two rates are equal: $g_{S}=g_{A}$.
The model is illustrated in Fig. \ref{fig:SAN_model}. We solve the
SAN-model analytically in the continuum limit and show how, at long
times, $3/2$-power-law distributions of cell populations asymptotically
arise near criticality. Fitting the model predictions to the empirical
data of Ref. \citep{esk_2020_organoid_tissue_screen}, we show that
cerebral organoid growth is indeed critical (see Fig. \ref{fig:theory_data_powerlaws_avalanches}
and Table \ref{tab:parameter_table}).

In experiments, organoids are grown for forty days. As we will demonstrate,
our model can describe tissue growth as a dynamical process with a
limited number of parameters -- three rates and one initial condition.

\paragraph*{Fokker-Planck description of lineage dynamics\label{sec:Fokker-Planck-description-of_lin_dyn}}

We start with a master equation for the SAN process. Individual populations
of S, A and N-cells are not accessible in experiments, since all descendants
of a stem cell inherit the same lineage identifier. We therefore need
to calculate the probability distribution of the total lineage size
$x_{\mathrm{tot}}=s+a+n$. This calls for a slight simplification.
We replace the processes A$\rightarrow$A + N and A$\rightarrow$N
by the assumption that each A-cell produces $\bar{N}=1+g_{N}/g_{F}$
N-cells over the course of its existence (where the $+1$ stems from
the final A$\rightarrow$N conversion) (Fig. \ref{fig:SAN_model}).
Since the N-cell output per A-cell varies by multiple orders of magnitude
less than the total offspring of an S-cell \citep{pflug2021_SAN},
this simplification does not affect the model's main predictions,
making it analytically tractable. $\bar{N}$ is a fitting parameter
in our theory. The total lineage size becomes $x_{\mathrm{tot}}=s+n$.
The master equation for the probability distribution $f$ of S- and
N-cell numbers $s$, $n$ at time $t$ then reads
\begin{eqnarray}
\partial_{t}f\left(s,n,t\right) & = & g_{S}\left(s-1\right)f\left(s-1,n,t\right)\nonumber \\
 & + & g_{A}\left(s+1\right)f\left(s+1,n-\bar{N},t\right)\nonumber \\
 & + & g_{0}\left(s+1\right)f\left(s+1,n,t\right)\nonumber \\
 & - & \left(g_{S}s+g_{A}s+g_{0}s\right)f\left(s,n,t\right).\label{eq:Master_eq}
\end{eqnarray}
The right hand side terms of Eq. (\ref{eq:Master_eq}) correspond
to the different processes that the cells undergo: S-cell death at
rate $g_{0}$, S-cell division at rate $g_{S}$ and differentiation
into $\bar{N}$ N-cells at rate $g_{A}$. Focusing on large cell counts,
we translate the discrete process of Eq. (\ref{eq:Master_eq}) into
a continuous version given by the Fokker-Planck equation \citep{vanKampen}:
\begin{equation}
\partial_{t}f\left(\mathbf{x},t\right)=\mathcal{L}f\left(\mathbf{x},t\right)\label{eq:Fokker_Planck_operator}
\end{equation}
with the differential operator 
\begin{align}
\mathcal{L} & =\left(-\alpha\partial_{s}+\frac{\beta}{2}\partial_{s}^{2}-g_{A}\bar{N}\partial_{s}\partial_{n}-\bar{N}g_{A}\partial_{n}+\frac{g_{A}\bar{N}^{2}}{2}\partial_{n}^{2}\right)s.\label{eq:Fokker-Planck}
\end{align}
Here, $\mathbf{x}=\left(s,n\right)$ is a vector with continuous cell
numbers as components, $\alpha=g_{S}-g_{A}-g_{0}$ and $\beta=g_{S}+g_{A}+g_{0}$.

\paragraph*{Regimes of growth: power-laws and avalanches\label{sec:Tissue-growth:-power-laws}}

Next, we want to examine the implications of the Fokker-Planck Eq.
(\ref{eq:Fokker_Planck_operator}) for the lineage sizes within a
tissue sample. We solve Eq. (\ref{eq:Fokker_Planck_operator}) with
the initial condition $f\left(\mathbf{x},t=0\right)=\delta\left(s-s_{0}\right)\delta\left(n\right)$.
$s_{0}$ corresponds to the initial number of stem cells of a lineage
-- not necessarily unity, since stem cells proliferate at initial
stages of organoid preparation, which are not considered here otherwise.
Using the Fourier transform of Eqs. (\ref{eq:Fokker_Planck_operator}),
(\ref{eq:Fokker-Planck}) with respect to $\mathbf{x}$ and the method
of characteristics to solve the resulting first order partial differential
equation (see supplementary material), we find the characteristic
function of $f\left(\mathbf{x},t\right)$, $\tilde{f}\left(\mathbf{q},t\right)=\int_{-\infty}^{\infty}e^{-i\mathbf{q}\cdot\mathbf{x}}f\left(\mathbf{x},t\right)$.

The distribution of total lineage sizes $x_{\mathrm{tot}}$ is given
by an integral of $f\left(\mathbf{x},t\right)$ over all states with
equal $x_{\mathrm{tot}}$:
\begin{equation}
f_{\mathrm{tot}}\left(x_{\mathrm{tot}},t\right)=\int_{0}^{x_{\mathrm{tot}}}f\left(s,x_{\mathrm{tot}}-s,t\right)ds.\label{eq:x_tot_distribution_def}
\end{equation}

The presence of a critical point can be clearly seen from the expectation
value of the stem cell population: for finite $\alpha$, $\left\langle s\right\rangle =s_{0}e^{\alpha t}$,
whereas for the critical $\alpha=0$, $\left\langle s\right\rangle =s_{0}$.
For $t\rightarrow\infty$, $f_{\mathrm{tot}}\left(x_{\mathrm{tot}},t\right)$
approaches a limiting distribution $f_{\mathrm{tot}}^{t\rightarrow\infty}\left(x_{\mathrm{tot}}\right)$
-- a truncated $3/2$-power-law Lévy distribution \citep{tsallis1997levy,gnedenko1954_stable_distr,bouchaud1990anomalous}:
\begin{equation}
f_{\mathrm{tot}}^{t\rightarrow\infty}\left(x_{\mathrm{tot}}\right)\approx\frac{s_{0}\beta e^{-\frac{\alpha^{2}x_{\mathrm{tot}}}{\beta^{2}\bar{N}}}}{2\sqrt{2\pi}\bar{N}\left(x_{\mathrm{tot}}/\bar{N}\right)^{3/2}}.\label{eq:critical_Levy_distr}
\end{equation}
This formula holds for $\alpha\ll\beta$. For a more general expression
see supplemetary Eq. (S 28). At criticality the distribution becomes
a true $3/2$-power-law.

The way in which $f_{\mathrm{tot}}\left(x_{\mathrm{tot}},t\right)$
approaches the limit of Eq. (\ref{eq:critical_Levy_distr}) is very
different for $\alpha>0$ and $\alpha<0$. For positive $\alpha$
and large enough lineage size $x_{\mathrm{tot}}>x_{\mathrm{tot}}^{*}\sim e^{\alpha t}$,
there is a region where $f_{\mathrm{tot}}\left(x_{\mathrm{tot}},t\right)$
is not approximated by Eq. (\ref{eq:critical_Levy_distr}). This is
the \textit{avalanche region} illustrated in Fig. \ref{fig:theory_data_powerlaws_avalanches}
c). Here, lineages have a high percentage of proliferating S-cells
which are driving the system's growth. The lineage sizes are very
large but the probability of their occurrence is very small. For $x_{\mathrm{tot}}<x_{\mathrm{tot}}^{*}$
most lineages are fully differentiated and stopped growing. In the
avalanche regime, $f_{\mathrm{tot}}\left(x_{\mathrm{tot}},t\right)$
can be approximated by 
\begin{equation}
f_{\mathrm{tot}}^{\mathrm{av.}}\left(x_{\mathrm{tot}},t\right)\approx\frac{\sqrt{a}\exp\left(-\frac{a}{q_{\mathrm{tot}}^{*}}-\frac{x_{\mathrm{tot}}}{q_{\mathrm{tot}}^{*}N}\right)}{q_{\mathrm{tot}}^{*}\sqrt{x_{\mathrm{tot}}\bar{N}}}I_{1}\left(2\sqrt{\frac{ax_{\mathrm{tot}}}{q_{\mathrm{tot}}^{*2}\bar{N}}}\right)\label{eq:avalanche_approx}
\end{equation}
where $I_{1}\left(z\right)$ is the modified Bessel function of the
first kind and $q_{\mathrm{tot}}^{*}\left(t\right)$ and $a\left(t\right)$
are defined in supplementary Eqs. (S 37) and (S 42).

For $\alpha<0$, we also find an avalanche region with an active S-cell
population approximately described by Eq. (\ref{eq:avalanche_approx})
(even though the avalanche will stop eventually). As pointed out in
the supplement, the approximation breaks down for $\alpha t>1$, but
is still reasonable for the data at hand. This region is located at
large $x_{\mathrm{tot}}$, for which $f_{\mathrm{tot}}\left(x_{\mathrm{tot}},t\right)>f_{\mathrm{tot}}^{t\rightarrow\infty}\left(x_{\mathrm{tot}}\right)$,
followed by a rapid truncation at even larger $x_{\mathrm{tot}}$
(Fig. \ref{fig:theory_data_powerlaws_avalanches} b)). In contrast
to the behavior at $\alpha>0$, $f_{\mathrm{tot}}\left(x_{\mathrm{tot}},t\right)$
convergences to $f_{\mathrm{tot}}^{t\rightarrow\infty}\left(x_{\mathrm{tot}}\right)$
uniformly: the avalanches becomes less and less pronounced as $t\rightarrow\infty$,
because the S-cells of all lineages eventually fully differentiate
if $\alpha\leq0$ \citep{Feller_Kampf_ums_Dasein}. Our analysis of
the data shows that $\alpha\lesssim0$ holds in experiments (see Table
\ref{tab:parameter_table}).

\paragraph*{Dynamics and criticality in experiments}

The experimental data consists of lineage identifier counts for 15
organoids that were sequenced on days 16, 21, 25, 32 and 40 -- three
copies for each day\citep{esk_2020_organoid_tissue_screen}. Accounting
for statistical and readout errors, they can be related to lineage
sizes \citep{pflug2021_SAN}. Each organoid consists of $\sim10^{4}$
lineages, while total cell counts evolve from $\sim10^{5}$ at day
16 to $\sim10^{6}$ at day 40. The organoids grow undisturbed from
day 11 onward, when they are not subjected to intrusive procedures
anymore. Four parameters are determined from experimental data: $\alpha$,
$\beta$, $s_{0}$ and $\bar{N}$. The data at day 40 is very roughly
distributed according to an $x^{-3/2}$ power-law (see Fig. \ref{fig:theory_data_powerlaws_avalanches}),
indicating near critical growth with $\alpha\ll\beta$. 

To determine the parameters precisely, we use the empirical characteristic
function of the data
\begin{equation}
\tilde{f}_{\mathrm{exp}}\left(q_{\mathrm{tot}}\right)=\sum_{k}e^{iq_{\mathrm{tot}}x_{\mathrm{tot}}^{\left(k\right)}}.\label{eq:empirical_characteristic}
\end{equation}
Here $x_{\mathrm{tot}}^{\left(i\right)}$ are the experimentally determined
lineage sizes. The characteristic function of our model $\tilde{f}\left(q_{\mathrm{tot}},t\right)$
(Eq. (S 33)) is then least squares fitted to $\tilde{f}_{\mathrm{exp}}\left(q_{\mathrm{tot}}\right)$.
Moving the fitting procedure to Fourier space has several advantages:
$\tilde{f}_{\mathrm{exp}}\left(q_{\mathrm{tot}}\right)$ is less noisy
than the empirical probability distribution and no smoothing procedures
such as kernel density estimates need to be used. Approximations needed
when transforming $\tilde{f}\left(q_{\mathrm{tot}},t\right)$ to real
space are avoided. Model fitting via the characteristic function has
been considered e.g. in Refs. \citep{yu2004_characteristic_function_fitting,chan2009_characteristic_function_fitting}.
Our estimates for the model parameters are shown in Table \ref{tab:parameter_table}.
\begin{table}
\centering{}%
\begin{tabular}{|c|c|}
\hline 
$\alpha\left[\mathrm{day}^{-1}\right]$ & $-0.018\pm0.004$\tabularnewline
\hline 
$\beta\left[\mathrm{day}^{-1}\right]$ & $4\pm1$\tabularnewline
\hline 
$s_{0}$ & $9\pm3$\tabularnewline
\hline 
$\bar{N}$ & $2\pm1$\tabularnewline
\hline 
\end{tabular}\caption{Parameters of the SAN model estimated from experimental data with
one standard deviation errors. $\alpha$ is the net growth rate of
the stem cell population, and $\beta$ characterizes the stochasticity
of the system. $s_{0}$ and $\bar{N}$ are the average initial number
of stem cells and the average number of N-cells produced by each stem
cell, respectively. \label{tab:parameter_table}}
\end{table}
In Fig. \ref{fig:theory_data_powerlaws_avalanches} a) we show a comparison
between the probability densities of the SAN model $f_{\mathrm{tot}}\left(x_{\mathrm{tot}},t\right)$
and binned histograms of the data at days 16 and 40. We used 500 bins
for the data at day 40, and 80 bins for day 16. In the supplement,
we show that the agreement is independent on bin-size. Small lineages
($<20$ cells) were excluded since these lineages, with a high probability,
died out at preparatory stages. $f_{\mathrm{tot}}\left(x_{\mathrm{tot}},t\right)$
is found using a Fast Fourier Transform of $\tilde{f}\left(q_{\mathrm{tot}},t\right)$.

The estimates of Table \ref{tab:parameter_table} show that organoid
growth is indeed a critical process with $\left|\alpha\right|\ll\beta$.
It also shows that even at day $40$ the process is far from its $t\rightarrow\infty$
limit ($\alpha t\approx0.8$) and the organoids maintain an avalanche-like
population of S-cells at large $x_{\mathrm{tot}}$.

The \textit{SAN model} is only a minimal model of biological reality
and as such cannot account for the full spread of the experimental
data. It does, for example, not account for the data at small $x_{\mathrm{tot}}$
(see Fig. \ref{fig:theory_data_powerlaws_avalanches}). Many small
lineages have died out in the early stages of the organoid development
and are obviously not described by SAN dynamics. We assumed that all
lineages consist of $s_{0}$ stem cells at day 11. This is only an
average. While for large lineages the initial number of stem cells
is unimportant since the growth is stochastic, it matters for lineages
that differentiated quickly. Other neglected aspects are the time
dependence of rates, and the non-markovian nature of division and
differentiation. Despite these caveats, the \textit{SAN model} proves
to be surprisingly robust and fits the experimental data well. In
supplementary section D, we perform Kolmogorov-Smirnov (K-S) tests
on different intervals for the data at day 40. K-S tests are a sensitive
tool to test the power-law behavior of empirical data \citep{clauset2009power-law_Kolmogorov_Smirnov,touboul2010power_law_neural_avalanches_Kolmogorov_Smirnov,touboul2017power_law_neural_avalanches_Kolmogorov_Smirnov}.
We find that the test produces high p-values (up to $p\approx0.8$)
in the region between $x_{\mathrm{tot}}\approx200$ and $x_{\mathrm{tot}}\approx2000$,
where the power-law is most pronounced.

\paragraph*{Extinction trajectories}

\begin{figure*}
\includegraphics[width=0.9\textwidth]{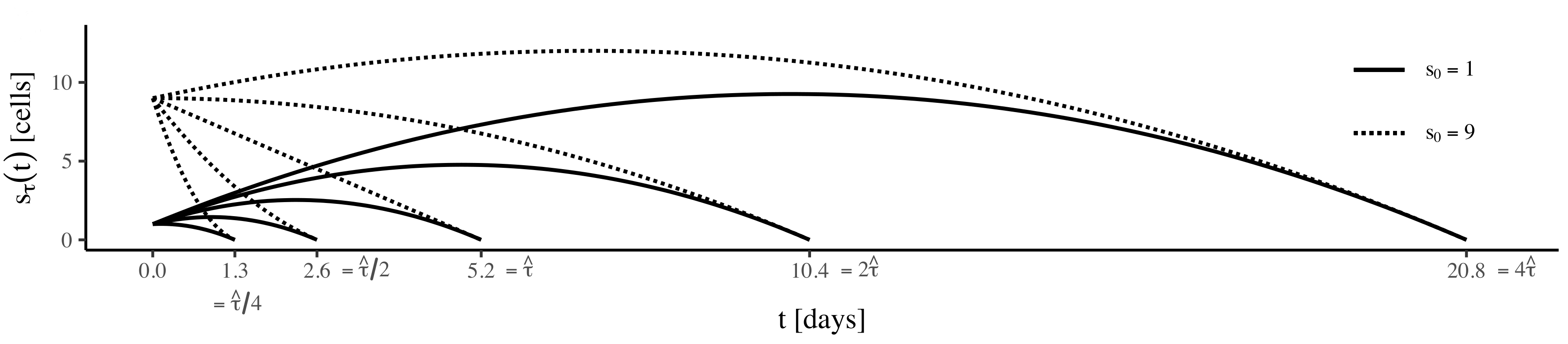}

\caption{Single lineage stem cell populations for initial sizes $s_{0}=1$
(solid lines) and $s_{0}=9$ (dotted lines) and different extinction
times $\tau$ (see Eq. (\ref{eq:extinct_traj})). For large extinction
times $\tau$, the trajectories for $s_{0}=1$ and $s_{0}=9$ become
equal for large times: In the critical state the population dynamics
is dominated by stochasticity, and hence obtains a universal form
independent of the initial conditions. \label{fig:extinction-trajectory}}
\end{figure*}

We now turn to the influence of criticality on the dynamics of single
lineages. We focus on $\alpha=0$ and restrict ourselves to S-cells
as the only dynamical component. Neglecting the dynamics of the $n$
variable, we drop all but the first two terms in $\mathcal{L}$ in
Eq. (\ref{eq:Fokker-Planck}). The reduced Fokker-Planck equation
is equivalent to the stochastic differential equation
\begin{equation}
ds=\sqrt{{\beta s}}dw(t).\label{eq:sde}
\end{equation}
$w\left(t\right)$ is the standard Brownian motion. It is known that
any $s\left(t\right)$ described by Eq. (\ref{eq:sde}), at some time
$\tau$, reaches $s\left(\tau\right)=0$ \citep{Feller_Kampf_ums_Dasein},
i.e. the S-cell population of the lineage goes extinct due to differentiation.
Using the Onsager-Machlup formalism \citep{durr_onsager-machlup_1978},
we find the \textit{extinction trajectory} 
\begin{equation}
s_{\tau}\left(t\right)=s_{0}\left(1-\frac{1}{\tau}\right)\left(1+\left(\frac{\tau}{\hat{\tau}}-1\right)\frac{t}{\tau}\right).\label{eq:extinct_traj}
\end{equation}
It describes the most probable path between $s\left(0\right)=s_{0}$
and $s\left(\tau\right)=0$ (see Fig. \ref{fig:extinction-trajectory}).
Here, $\hat{\tau}=4s_{0}/\left(\sqrt{3}\beta\right)$. Details of
the calculation are given in the supplementary material. Eq. (\ref{eq:extinct_traj})
and Fig \ref{fig:extinction-trajectory} show that at criticality,
the cells loose memory of the initial condition $s_{0}$, since for
$\hat{\tau}\ll t<\tau$
\begin{equation}
s_{\tau}\left(t\right)\sim\sqrt{3}\beta t\left(1-t/\tau\right)/4.\label{eq:univ_traj}
\end{equation}
This is not the case for $\alpha\gtrless0$.

Most importantly, since the lineage sizes are power-law distributed,
an overwhelming majority of organoid cells will belong to a few very
large lineages. For these lineages, $\tau\gg\hat{\tau}$ will hold,
meaning that their development will be largely independent of the
initial condition $s_{0}$ (Eq. (\ref{eq:univ_traj})). This is an
essential feature of critical growth: it is independent of the initial
conditions.

\paragraph*{Discussion}

Finally, we discuss possible implications of critical tissue growth.
Tuning itself to the critical point, the organoid maximizes the stochasticity
of the growth process. We are not dealing with stochastic growth at
a given rate. Instead, the characteristic time scale of growth vanishes
($\alpha=0$) and growth is fully determined by the stochasticity
of the process. This is in contrast to many other examples of organ
growth and regeneration \citep{fausto2000liver,zeng2008_organ_size_control,Ainslie2020_Drosophila_tissue_growth_kinetics}
where the growth process is arrested in a coordinated manner. In critical
growth, the long term dynamics of large lineages, which make up most
of the organoid, does not depend on initial conditions. Stochastic
fluctuations have erased all memory of the lineage's initial size.
This might hint at a biological advantage of the critical regime:
the outcome of the growth process is less influenced by perturbations
in its initial stages, when the tissue is most susceptible to disturbances.

Second, the critical nature of the process implies a mechanism balancing
the rates of division and differentiation of stem cells. Such balancing
mechanisms are known from homeostatic stem cell renewal \citep{simons_strategies_2011}.
One distinguishes between mechanisms based on asymmetric division
(only one daughter cell is a stem cell while the other differentiates)
and population asymmetry (the rates of stem cell loss and division
are balanced on population level). The first strategy cannot produce
critical lineage dynamics because it is strictly deterministic. On
the population level, stem cell niches are a well known regulatory
mechanism for homeostatic tissues \citep{moore2006stem,lin_classic_2015,Corominas2020_intestinal_niches}
found in intestinal crypts \citep{moore2006stem} and the adult human
brain \citep{alvarez2004long}. The available niche space limits the
number of possible divisions. For growing tissues, the competition
for a scarce resource, e.g. space or nutrients, can provide a feedback
loop that limits the stem cells' ability for division at the population
level. Such a feedback loop is often encountered in models of self
organized criticality (SOC) \citep{Zapperi1995_SOC}, where the event
probabilities are balanced and tuned to the critical point e.g. by
energy conservation. For cerebral tissue, the currently available
data gives no clues to the nature of the balancing mechanism. The
search, however, could inspire future experimental research. Recent
advances in lineage tracing allow to study the spatial distribution
of lineages in cerebral organoids using light-sheet microscopy and
spatial transcriptome sequencing \citep{he2022spatial_lineage_tracing}.
Inferring the spatial dynamics of the growth process could help to
identify the the mechanism behind organoid SOC. Quantitative lineage
tracing experiments with other organoid types such as intestinal \citep{angus2020intestinal_organoids,date2015intestinal_organoid},
retinal \citep{volkner2016retinal,quadrato2017cell} or cardiac organoids
\citep{nugraha2019human,hoang2018cardiac_organoids} could reveal
whether critical growth is specific to cerebral tissue, or whether
it is a more general organizing principle.
\begin{acknowledgments}
We thank C. Esk, N. Goldenfeld, S. Haendeler, B. Jeevanesan, J. F.
Karcher, and D. Lindenhofer for inspiring discussions. This project
has received funding from the European Union’s Framework Programme
for Research and Innovation Horizon 2020 (2014‐2020) under the Marie
Curie Skłodowska Grant Agreement Nr. 847548 (AvH, EK), and from a
Special Research Programme (SFB) of the Austrian Science Fund (FWF),
project number F78 P11 (AvH, FP).
\end{acknowledgments}

\clearpage\onecolumngrid

\setcounter{page}{1}
\section*{Supplementary Material} 

\renewcommand{\thesection}{Supplementary Sec.}%

\setcounter{figure}{0}
\renewcommand{\thefigure}{S\ \arabic{figure}}%
\renewcommand{\figurename}{Supplementary Figure}

\setcounter{equation}{0}
\renewcommand{\theequation}{S\,\arabic{equation}}%

\subsection{Solving the Fokker-Planck equation\label{sec:Solving-the-Fokker-Planck}}

All supplementary equations are marked with an ``S'', while equation
numbers without ``S'' refer to the main text. The idea that a laplace
transform of the S-cell (branching) part of the Fokker-Planck equation
(\ref{eq:Fokker_Planck_operator}) yields a solvable first order partial
differential equation was used by Feller in Ref. \citep{feller1951_Two_singular_diffusion_problems}.
Here we use the Fourier transform to obtain the distributions of total
population sizes of the branching based SAN-model near criticality.

Although cell counts are discrete, we chose a continuous model to
describe the data. For small $s$ the results of the continuous and
discrete approaches will differ, yet both are reasonable approximations
of biological reality. While using the discrete version may seem preferential
at first, cell division is only the end result of a series of changes
to a cell's internal state as it transitions through the G1, S, G2
and M stages of the cell cycle. Because the continuous process accounts
for these changes by allowing cell counts to change gradually, it
may be in fact the model that is closer to reality.

\subsubsection{The characteristic function}

The Fokker-Planck equation (\ref{eq:Fokker_Planck_operator}) can
be solved in the Fourier domain. After a Fourier transform, Eq. (\ref{eq:Fokker_Planck_operator})
becomes
\begin{eqnarray}
\frac{\partial\tilde{f}\left(\mathbf{q},t\right)}{\partial t} & = & -\left[\frac{\beta}{2}q_{s}^{2}+\alpha iq_{s}+ig_{A}\left(\bar{N}q_{n}\right)+\frac{g_{A}}{2}\left(\bar{N}q_{n}\right)^{2}-g_{A}q_{s}\left(\bar{N}q_{n}\right)\right]\left(i\frac{\partial}{\partial q_{s}}\right)\tilde{f}\left(\mathbf{q},t\right)\label{eq:Fokker-Planck-Fourier}
\end{eqnarray}
 with
\begin{equation}
\tilde{f}\left(\mathbf{q},t\right)=\int d^{3}x\,e^{-i\mathbf{q}\cdot\mathbf{x}}f\left(\mathbf{x},t\right),
\end{equation}
where $\mathbf{q}=\left(s,n\right)$. In the following we will drop
the factors of $\bar{N}$ for notational simplicity and restore them
later on with the substitution 
\begin{equation}
q_{n}\rightarrow\bar{N}q_{n}.\label{eq:N_N_cells_substitiution}
\end{equation}
Let us choose the initial condition
\begin{equation}
f\left(\mathbf{x},t=0\right)=\delta\left(s_{0}-s\right)\delta\left(n\right),
\end{equation}
which corresponds to $s_{0}$ S-cells at $t=0$. In Fourier space
this initial condition translates to
\begin{equation}
\tilde{f}\left(\mathbf{q},t=0\right)=e^{-is_{0}q_{s}}.\label{eq:Fourier_initial}
\end{equation}
Eq. (\ref{eq:Fokker-Planck-Fourier}) contains only first order derivatives
and can be solved with the method of characteristics. The characteristic
equations are
\begin{align}
\frac{\partial t}{\partial\tau} & =1\nonumber \\
\frac{\partial q_{s}}{\partial\tau} & =\left[-\alpha q_{s}+i\frac{\beta}{2}q_{s}^{2}-g_{A}q_{n}+i\frac{g_{A}}{2}q_{n}^{2}-ig_{A}q_{s}q_{n}\right].
\end{align}
The first equation is trivial. It is solved by $t=\tau$. The second
equation is solved by
\begin{equation}
q_{s}\left(t\right)=\frac{1}{\beta}\frac{i\kappa\left(q_{n}\right)\left(C_{1}e^{t\kappa\left(q_{n}\right)}-1\right)}{C_{1}e^{t\kappa\left(q_{n}\right)}+1}+\frac{g_{A}q_{n}-i\alpha}{\beta}
\end{equation}
with
\begin{align}
\kappa\left(q_{n}\right) & =\sqrt{-\left(g_{A}q_{n}-i\alpha\right)^{2}-2i\beta D\left(q_{n}\right)}\\
D\left(q_{n}\right) & =-g_{A}q_{n}+i\frac{g_{A}}{2}q_{n}^{2}.
\end{align}
The solution for the Fourier transformed Fokker-Planck equation (\ref{eq:Fokker-Planck-Fourier})
is found by solving for the integration constant $C_{1}$:
\begin{equation}
C_{1}=\frac{\alpha+\kappa\left(q_{n}\right)+ig_{A}q_{n}-i\beta q_{s}}{\kappa\left(q_{n}\right)-\alpha-ig_{A}q_{n}+i\beta q_{s}}e^{-t\kappa\left(q_{a}\right)}.
\end{equation}
The next step is to look for a function of $C_{1}$, $G\left(C_{1}\left(q_{s},q_{n},t\right)\right)$,
which satisfies
\begin{equation}
\left.G\left(C_{1}\right)\right|_{t=0}=q_{s},
\end{equation}
so that a function $\tilde{f}\left(\mathbf{q},t\right)$ that satisfies
the initial condition of Eq. (\ref{eq:Fourier_initial}) can be written
as
\begin{equation}
\tilde{f}\left(\mathbf{q},t\right)=e^{-is_{0}G\left(C_{1}\right)}.\label{eq:charact_funct_full}
\end{equation}
Such a function is given by
\begin{align}
G\left(q_{s},q_{n},t\right) & =\frac{q_{s}\rho\left(q_{n},t\right)-\frac{2}{\beta}\lambda\left(q_{n},t\right)D\left(q_{a}\right)}{1+i\lambda\left(q_{n},t\right)q_{s}}\label{eq:Fourier_G_Solution_qn=00003D0}\\
\lambda\left(q_{n},t\right) & =\frac{\beta}{\kappa\left(q_{n}\right)\coth\left(\frac{1}{2}\kappa\left(q_{n}\right)t\right)-\alpha-ig_{A}q_{n}}\nonumber \\
\rho\left(q_{n},t\right) & =\frac{\kappa\left(q_{n}\right)\coth\left(\frac{1}{2}\kappa\left(q_{n}\right)t\right)+\alpha+ig_{A}q_{n}}{\kappa\left(q_{n}\right)\coth\left(\frac{1}{2}\kappa\left(q_{n}\right)t\right)-\alpha-ig_{A}q_{n}}.\nonumber 
\end{align}
The function $\tilde{f}\left(\mathbf{q},t\right)$ of Eq. (\ref{eq:charact_funct_full})
is the characteristic function of $f\left(\mathbf{x},t\right)$. For
later purposes we want to separate $\kappa\left(q_{n}\right)$ into
real and imaginary parts. We find
\begin{align}
\mathrm{Re}\left(\kappa\left(q_{n}\right)\right) & =\sqrt{\frac{\sqrt{4g_{A}^{2}q_{n}^{2}\left(\alpha+\beta\right)^{2}+\left(\alpha^{2}+g_{A}q_{n}^{2}\left(\beta-g_{A}\right)\right)^{2}}-g_{A}^{2}q_{n}^{2}+\beta g_{A}q_{n}^{2}+\alpha^{2}}{2}}\label{eq:kappa_real}\\
\mathrm{Im}\left(\kappa\left(q_{n}\right)\right) & =\text{sign}\left(q_{n}\right)\sqrt{\frac{\sqrt{4g_{A}^{2}q_{n}^{2}\left(\alpha+\beta\right)^{2}+\left(\alpha^{2}+g_{A}q_{n}^{2}\left(\beta-g_{A}\right)\right)^{2}}+g_{A}^{2}q_{n}^{2}-\beta g_{A}q_{n}^{2}-\alpha^{2}}{2}}.\label{eq:kappa_imaginary}
\end{align}

\subsubsection{Large $t$ limit. Distribution of N cells.\label{subsec:Large-t-limit.}}

Let us gain some intuition into the dynamics of the growth process.
If the S-cells are dividing at a near critical near zero rate $\alpha$,
while they are differentiating at a much larger rate $g_{A}$, we
can expect that, as $t$ grows, most lineages will consist of differentiated
cells. Lineages that escaped differentiation will have to be very
lucky and sufficiently large. As a first step, we want to consider
the distribution differentiated cells.

\begin{figure}
\includegraphics[scale=0.4]{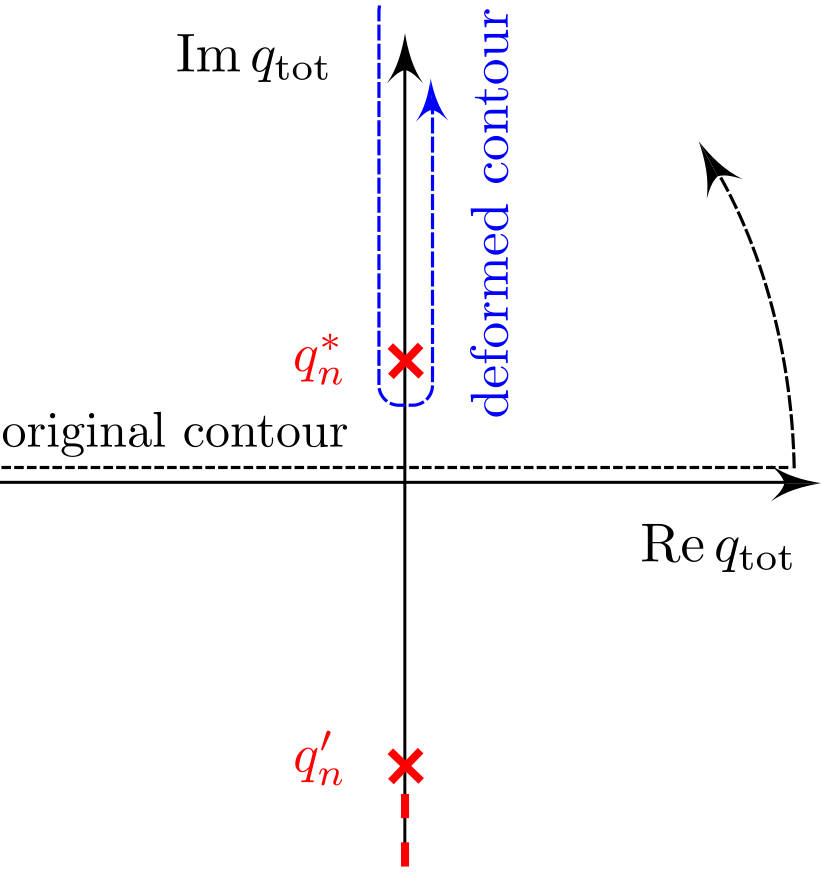}\caption{The branch-cut and contour integration in Eq. (\ref{eq:xs=00003D0_new_contour}).
\label{fig:Branch-cut-integration}}
\end{figure}

The inverse Fourier transform of Eq. (\ref{eq:charact_funct_full})
with respect to $q_{s}$ reads
\begin{align}
f\left(s,q_{a},t\right) & =\int\frac{dq_{s}}{2\pi}e^{iq_{s}s-is_{0}G\left(q_{s},q_{n},t\right)}.
\end{align}
This expression can be rewritten as
\begin{align}
f\left(s,q_{n},t\right) & =e^{-\frac{\rho s_{0}}{\lambda}}\int\frac{dq_{s}}{2\pi}e^{iq_{s}s}\left[\exp\left(\frac{is_{0}2D\left(q_{n}\right)\lambda/\beta+\rho s_{0}/\lambda}{1+i\lambda q_{s}}\right)-1+1\right]\nonumber \\
 & =e^{-\frac{\rho s_{0}}{\lambda}}\int\frac{dq_{s}}{2\pi}e^{iq_{s}s}\left[\exp\left(\frac{is_{0}2D\left(q_{n}\right)\lambda/\beta+\rho s_{0}/\lambda}{1+i\lambda q_{s}}\right)-1\right]+e^{-\frac{\rho s_{0}}{\lambda}}\delta\left(x_{s}\right).\label{eq:charact_funct_delta_manipulation}
\end{align}
We have thus separated the function $f\left(x_{s,}q_{n},t\right)$
into a part which depends on $x_{s}$ and describes lineages that
are still evolving and a second part with $x_{s}=0$. This latter
part contains information on the distribution of N-cells in fully
differentiated lineages. Its inverse Fourier transform with respect
to $q_{n}$ is given by
\begin{align}
f\left(s=0,n\right) & =e^{-\frac{\alpha}{\beta}s_{0}}\int\frac{dq_{n}}{2\pi}\exp\left(iq_{n}\left(n-s_{0}\frac{g_{A}}{\beta}\right)-s_{0}\frac{\kappa\left(q_{n}\right)}{\beta}\right)\label{eq:fully_evolved_n_distr}
\end{align}
Notice that $f\left(s=0,n\right)$ does not depend on time. We will
later see that $f\left(s=0,n\right)$ is the $t\rightarrow\infty$
limit of the distribution of total lineage sizes (see Eq. (\ref{eq:x_tot_distribution_def})
in the main text). Note that the integrand of the first right hand
side term in the second line of Eq. (\ref{eq:charact_funct_delta_manipulation})
approaches zero as $q_{s}\rightarrow\infty$. This is a consequence
of separating out the delta function and is necessary to regularize
the integral. We will attend to this issue below. To do the integral
in Eq. (\ref{eq:fully_evolved_n_distr}), it is useful to take a look
at the analytic structure of the integrand. Nonanalyticities arise
from the square root structure of $\kappa\left(q_{n}\right)$. Two
branch cuts start at the two imaginary roots of $\kappa\left(q_{n}\right)$
and run to $+i\infty$ and $-i\infty$ along the imaginary $q_{n}$
axis (see Fig. \ref{fig:Branch-cut-integration}). The positive imaginary
root is 
\begin{equation}
iq_{n}^{*}\equiv\frac{i\alpha^{2}}{g_{A}\left(\alpha+\beta\right)+\sqrt{\beta g_{A}\left(\alpha^{2}+2\alpha g_{A}+\beta g_{A}\right)}},\label{eq:critical_upper_branch_cut_qn}
\end{equation}
while for the negative imaginary root we find
\begin{equation}
iq_{n}^{'}=\frac{i\alpha^{2}}{g_{A}\left(\alpha+\beta\right)-\sqrt{\beta g_{A}\left(\alpha^{2}+2\alpha g_{A}+\beta g_{A}\right)}}.
\end{equation}
For small $\left|\alpha\right|\ll\beta$, we have
\begin{align}
iq_{n}^{*} & \approx\frac{i\alpha^{2}}{2\beta g_{A}},\\
iq_{n}^{'} & \approx-\frac{2i\beta}{\beta-g_{A}}-\frac{2i\alpha}{\beta-g_{A}}.\label{eq:poles_approx}
\end{align}

We thus make a crucial observation: The branch-cut in the upper complex
half-plane of $q_{n}$ descends to the origin for $\alpha=0$, while
the branch cut in the lower half plane always starts below $-2i\beta/\left(\beta-g_{A}\right)$.
For $n>0$, the integral over the real line in the inverse Fourier
transform in Eq. (\ref{eq:fully_evolved_n_distr}) can be deformed
to an integral around the branch cut running from $+i\infty$ to $q_{a}^{*}$
to the left hand side of the branch cut (at a small distance $\varepsilon$,
say), and then running back to infinity on the right hand side (see
Fig. \ref{fig:Branch-cut-integration}). The half-circle around $q_{a}^{*}$
is of order $\varepsilon$ and can be neglected. We can write
\begin{align}
f\left(s=0,n\right) & =e^{-\frac{\alpha s_{0}}{\beta}}\int\frac{dq_{n}}{2\pi}\exp\left\{ iq_{n}\left(n-s_{0}\frac{g_{A}}{\beta}\right)-s_{0}\frac{\kappa\left(q_{n}\right)}{\beta}\right\} \nonumber \\
 & =e^{-\frac{\alpha s_{0}}{\beta}}\int_{\infty}^{q_{n}^{*}}\frac{idq_{n}}{2\pi}\exp\left\{ -\left(q_{n}-i\varepsilon\right)\left(n-s_{0}\frac{g_{A}}{\beta}\right)-s_{0}\frac{\kappa\left(iq_{n}+\varepsilon\right)}{\beta}\right\} \nonumber \\
 & \quad+e^{-\frac{\alpha s_{0}}{\beta}}\int_{q_{n}^{*}}^{\infty}\frac{idq_{n}}{2\pi}\exp\left\{ -\left(q_{n}+i\varepsilon\right)\left(n-s_{0}\frac{g_{A}}{\beta}\right)-s_{0}\frac{\kappa\left(iq_{n}-\varepsilon\right)}{\beta}\right\} ,\label{eq:xs=00003D0_new_contour}
\end{align}
On the new contour, we need
\begin{equation}
\kappa\left(iq_{n}\right)=\sqrt{\left(\alpha-g_{A}q_{n}\right)^{2}-\beta g_{A}q_{n}\left(q_{n}+2\right)}.
\end{equation}
It is clear from Eq. (\ref{eq:xs=00003D0_new_contour}) that the most
important contribution to the integrals will come from the vicinity
of $q_{n}^{*}$, because the integrand decays exponentially for larger
$q_{n}$ the faster, the larger $n$. Expanding $\kappa\left(iq_{n}\right)$
around $q_{n}^{*}$, we obtain
\begin{equation}
\kappa\left(iq_{n}\pm\varepsilon\right)=\pm i\sqrt{2\sqrt{\beta g_{A}\left(2\alpha g_{A}+\alpha^{2}+\beta g_{A}\right)}\left(q-q_{n}^{*}\right)}
\end{equation}
(approaching the branch cut from different sides changes the sign
of the square root). Thus, for large $n$, we approximate Eq. (\ref{eq:xs=00003D0_new_contour})
as
\begin{align}
f\left(s=0,n\right) & \approx-e^{-\frac{\alpha s_{0}}{\beta}}\int_{q_{n}^{*}}^{\infty}\frac{idq_{a}}{2\pi}\exp\left\{ -q_{n}\left(n-s_{0}\frac{g_{A}}{\beta}\right)-\frac{s_{0}}{\beta}i\sqrt{2\sqrt{\beta g_{A}\left(2\alpha g_{A}+\alpha^{2}+\beta g_{A}\right)}\left(q_{n}-q_{n}^{*}\right)}\right\} \nonumber \\
 & \quad+e^{-\frac{\alpha s_{0}}{\beta}}\int_{q_{n}^{*}}^{\infty}\frac{idq_{a}}{2\pi}\exp\left\{ -q_{n}\left(n-s_{0}\frac{g_{A}}{\beta}\right)+\frac{s_{0}}{\beta}i\sqrt{2\sqrt{\beta g_{A}\left(2\alpha g_{A}+\alpha^{2}+\beta g_{A}\right)}\left(q_{n}-q_{n}^{*}\right)}\right\} \nonumber \\
 & =-ie^{-\frac{\alpha s_{0}}{\beta}}\int_{0}^{\infty}\frac{dq_{a}}{2\pi}\exp\left\{ -\left(q_{n}+q_{n}^{*}\right)\left(n-s_{0}\frac{g_{A}}{\beta}\right)-\frac{s_{0}}{\beta}i\sqrt{2pq_{n}}\right\} \nonumber \\
 & \quad+ie^{-\frac{\alpha s_{0}}{\beta}}\int_{0}^{\infty}\frac{dq_{a}}{2\pi}\exp\left\{ -\left(q_{n}+q_{n}^{*}\right)\left(n-s_{0}\frac{g_{A}}{\beta}\right)+\frac{s_{0}}{\beta}i\sqrt{2pq_{n}}\right\} \nonumber \\
 & =2\mathrm{Im}\left(e^{-\frac{\alpha s_{0}}{\beta}}\int_{0}^{\infty}\frac{dq_{a}}{2\pi}\exp\left\{ -\left(q_{n}+q_{n}^{*}\right)\left(n-s_{0}\frac{g_{A}}{\beta}\right)+\frac{s_{0}}{\beta}i\sqrt{2pq_{n}}\right\} \right)
\end{align}
with
\begin{equation}
p=\sqrt{\beta g_{A}\left(2\alpha g_{A}+\alpha^{2}+\beta g_{A}\right)}
\end{equation}
For the remaining integral we obtain
\begin{align}
f\left(s=0,n\right) & \approx\frac{s_{0}\sqrt{p}e^{\frac{-ps_{0}^{2}}{2\beta^{2}\left(n-s_{0}\frac{g_{A}}{\beta}\right)}-q_{n}^{*}\left(n-s_{0}\frac{g_{A}}{\beta}\right)}}{\sqrt{2\pi}\beta\left(n-s_{0}\frac{g_{A}}{\beta}\right)^{3/2}}\nonumber \\
 & \approx\frac{s_{0}\sqrt{p}e^{-q_{n}^{*}\left(n-s_{0}\frac{g_{A}}{\beta}\right)}}{\sqrt{2\pi}\beta\left(n-s_{0}\frac{g_{A}}{\beta}\right)^{3/2}}\label{eq:evolved_lineages_powerlaw}
\end{align}
This is a truncated, one-sided Lévy distribution \citep{tsallis1997levy,gnedenko1954_stable_distr,bouchaud1990anomalous}.
The appearance of the Lévy distribution could have been anticipated,
since the integrand of Eq. (\ref{eq:fully_evolved_n_distr}) for $\alpha=0$
can be mapped to the characteristic function of the Lévy distribution
after the argument of the square root in $\kappa\left(q_{n}\right)$
has been expanded for small $q_{n}$. For small $\alpha$ and $g_{0}$
we find $g_{A}=\left(-\alpha+\beta-g_{0}\right)/2\approx\beta/2$,
as well as $p\approx\beta^{2}/2$, and for $n\gg1$ the truncated
power law of Eq. (\ref{eq:evolved_lineages_powerlaw}) becomes Eq.
(\ref{eq:critical_Levy_distr}) of the main text with $\bar{N}=1$.
In the main text, we stated that the $t\rightarrow\infty$ limit of
$f_{\mathrm{tot}}\left(x_{\mathrm{tot}},t\right)$ -- the distribution
of total lineage sizes $x_{\mathrm{tot}}$ -- is governed by the
same expression as $f\left(s=0,n\right)$ (see Eq. (\ref{eq:critical_Levy_distr})
of the main text). We will justify this statement below. On an intuitive
level this can be understood as a consequence of the fact that, as
$t$ increases, most cells will already have differentiated and $f_{\mathrm{tot}}$
will be dominated by the N-cell distribution.

Since $q_{n}^{*}$ descends to the origin as $\alpha\rightarrow0$,
the truncation of the $3/2$-power-law of (\ref{eq:evolved_lineages_powerlaw})
vanishes and (\ref{eq:evolved_lineages_powerlaw}) becomes a true
Lévy stable power-law distribution. Returning to the observation that
the integrand of (\ref{eq:fully_evolved_n_distr}) is nonanalytic
in the lower complex half plane, we conclude that $f\left(s=0,n\right)$
is finite for $n<0$, meaning that there is a finite probability to
find a negative number of N-cells. This is a consequence of approximating
the discrete master equation by a continuous Fokker-Planck equation
which is accurate for large $s$ and $n$. However, one can easily
convince oneself that $f\left(s=0,n\right)$ decays very fast as $n$
decreases below zero: Following the reasoning of the above branch-cut
integration, we see that $f\left(s=0,n\right)$ will be suppressed
by an exponential factor $e^{-q_{n}'n}$. On the other hand, we see
from Eq. (\ref{eq:poles_approx}) that for small $\alpha$ 
\begin{equation}
\left|q_{n}^{'}\right|\gg\left|q_{n}^{*}\right|\label{eq:branch-cut_difference}
\end{equation}
holds. Thus we conclude that the probability for a negative $n$ is
very small, and can be neglected as an artifact of the Fokker-planck
approximation. Notice that the probability for negative $s$ is zero,
because Eq. (\ref{eq:charact_funct_full}) is analytic in $q_{s}$,
except for a single pole which is always in the upper complex half
plane (since $\mathrm{Re}\left(\lambda\left(q_{n}\right)>0\right)$
as Eqs. (\ref{eq:kappa_real}), (\ref{eq:Fourier_G_Solution_qn=00003D0})
indicate). Indeed, while the diffusion coefficient associated with
the second derivative in $s$ in Eq. (\ref{eq:Fokker-Planck}) vanishes
at $s=0$, the diffusion coefficient associated with $n$-diffusion
does not vanish at the origin and allows for a small leakage of probability
towards negative $n$.

So far we have been investigating the analytic structure of the $s=0$
contribution to $\tilde{f}\left(\mathbf{q},t\right)$. However similar
arguments carry over to the general case. Multiplying the numerator
and denominator of the function $G\left(q_{s},q_{n},t\right)$ in
Eq. (\ref{eq:Fourier_G_Solution_qn=00003D0}) by $\lambda\left(q_{n},t\right)^{-1}$
and keeping $q_{s}$ real for the moment, we see, that the characteristic
function $\tilde{f}\left(\mathbf{q},t\right)$ (Eq. (\ref{eq:charact_funct_full}))
is analytic in $q_{n}$ except for the branch-cuts of $\kappa\left(q_{n}\right)$
and the pole at $\lambda\left(q_{n},t\right)=i/q_{s}$. This pole
occurs when $\lambda\left(q_{n},t\right)$ is purely imaginary. $f\left(s,n,t\right)$
will decay for $n<0$ the faster, the lower the nonanalyticities with
$\mathrm{Im}\left(q_{n}\right)<0$ lie in the complex plane. Eq. (\ref{eq:branch-cut_difference})
shows that the contribution of the lower half branch-cut is indeed
negligible since it starts much further away from the origin than
the upper branch-cut. A numerical inspection of $\lambda\left(q_{n},t\right)$
shows that regions where $\lambda\left(q_{n},t\right)$ is purely
imaginary in the lower complex half plane of $q_{n}$ are indeed sufficiently
far from the origin for all reasonable parameter choices. We conclude
that the finite values of $f\left(s,n,t\right)$ for negative $n$
are simply an artifact of the Fokker-Planck approximation to the original
master equation (\ref{eq:Master_eq}) and are small.

\subsubsection{The lineage size distribution $f_{\mathrm{tot}}\left(x_{\mathrm{tot}},t\right)$}

As pointed out in the main text, we are ultimately interested in the
distribution of the lineage size 
\begin{equation}
x_{\mathrm{tot}}=s+n.\label{eq:lineage_size_appendix}
\end{equation}
This distribution is obtained by summing $f\left(s,n,t\right)$ over
all states which satisfy the condition (\ref{eq:lineage_size_appendix}):
\begin{equation}
f_{\mathrm{tot}}\left(x_{\mathrm{tot}},t\right)=\int_{0}^{x_{\mathrm{tot}}}f\left(s,x_{\mathrm{tot}}-s,t\right)ds\label{eq:def_total_lineage_distr_append}
\end{equation}
(see also Eq. (\ref{eq:x_tot_distribution_def}) in the main text).
Having argued in Sec. \ref{subsec:Large-t-limit.}, that the distribution
$f\left(s,n,t\right)$ corresponding to the characteristic function
(\ref{eq:charact_funct_full}) is confined to positive $s$ and --
to a good approximation -- to positive $n$, we can extend the integration
in Eq. (\ref{eq:def_total_lineage_distr_append}) to the complete
real axis:
\[
f\left(x_{\mathrm{tot}},t\right)\approx\int_{-\infty}^{\mathrm{\infty}}ds\,f\left(s,x_{\mathrm{tot}}-s,t\right).
\]
A Fourier transform then gives
\begin{equation}
\tilde{f}\left(q_{\mathrm{tot}},t\right)\approx\tilde{f}\left(q_{s}=q_{\mathrm{tot}},q_{n}=q_{\mathrm{tot}},t\right),\label{eq:charact_q->qtot}
\end{equation}
where the characteristic function of Eq. (\ref{eq:charact_funct_full})
is on the right hand side. This yields
\begin{equation}
\tilde{f}\left(q_{\mathrm{tot}},t\right)=\exp\left(\frac{q_{\mathrm{tot}}s_{0}\left[\alpha+2g_{A}+\kappa\left(q_{\mathrm{tot}}\right)\coth\left(\frac{1}{2}\kappa\left(q_{\mathrm{tot}}\right)t\right)\right]}{-i\alpha+g_{A}q_{\mathrm{tot}}-\beta q_{\mathrm{tot}}+i\kappa\left(q_{\mathrm{tot}}\right)\coth\left(\frac{1}{2}\kappa\left(q_{\mathrm{tot}}\right)t\right)}\right).\label{eq:total_charact_funct}
\end{equation}

\paragraph{Analytic expressions for the power-law and avalanche regimes}

\begin{figure}
\includegraphics[scale=0.3]{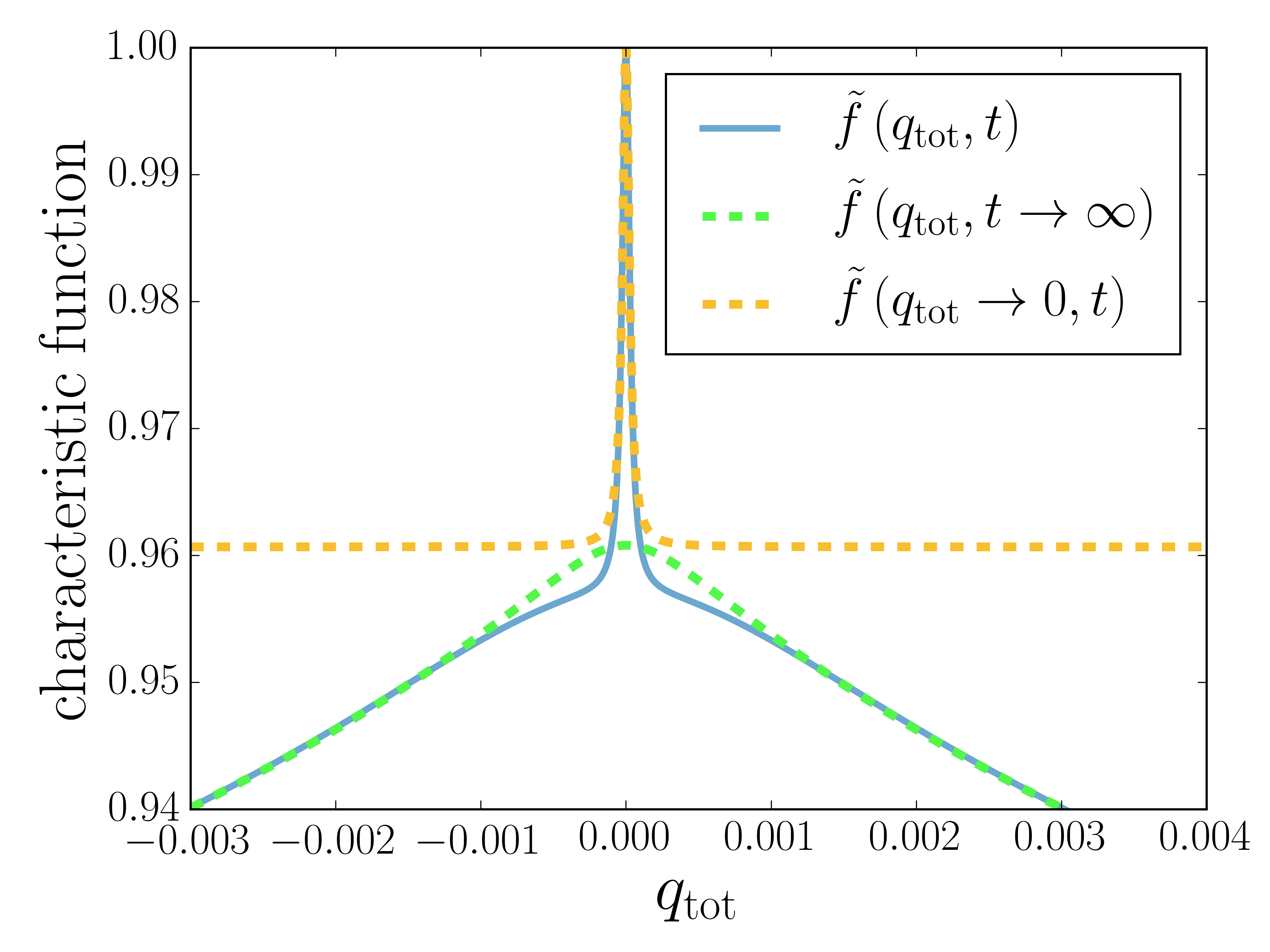}\caption{The full characteristic function $\tilde{f}\left(q_{\mathrm{tot}},t\right)$
of Eq. (\ref{eq:total_charact_funct}) and the approximations for
$t\rightarrow\infty$ or large $q_{\mathrm{tot}}$, $\tilde{f}\left(q_{\mathrm{tot}},t\rightarrow\infty\right)$,
(Eq. ($\ref{eq:tot_charact_t_inf}$\label{fig:charact_funct_approx})),
as well as for small $q_{\mathrm{tot}}$, $\tilde{f}\left(q_{\mathrm{tot}}\rightarrow0,t\right)$,
(Eq.\ref{eq:tot_charact_approx_small_q}). We used a positive growth
rate $\alpha=0.2/\mathrm{day}$ and $\beta=10/\mathrm{day}$. For
$\alpha>0$ and large but finite $t$, $\tilde{f}\left(q_{\mathrm{tot}},t\rightarrow\infty\right)$
(green dashed curve) is a good approximation to the characteristic
function $\tilde{f}\left(q_{\mathrm{tot}},t\right)$ (blue curve),
except for a small region near $q_{\mathrm{tot}}=0$ which is well
approximated by $\tilde{f}\left(q_{\mathrm{tot}}\rightarrow0,t\right)$
(orange curve). This region governs the behavior at large lineage
sizes $x_{\mathrm{tot}}$.}
\end{figure}

The characteristic function given in Eq. ($\ref{eq:total_charact_funct}$)
has a peculiar behavior at $q_{\mathrm{tot}}=0$ which determines
the asymptotics for large $x_{\mathrm{tot}}$. To see this, let us
first investigate the $t\rightarrow\infty$ and the $q_{\mathrm{tot}}\rightarrow\infty$
limits of $\tilde{f}\left(q_{\mathrm{tot}},t\right)$. Since $\mathrm{Re}\kappa\left(q_{\mathrm{tot}}\right)>0$
holds for all real $q_{\mathrm{tot}}$ (see Eq. (\ref{eq:kappa_real}))
and $\mathrm{Re}\kappa\left(q_{\mathrm{tot}}\right)\sim\left|q_{\mathrm{tot}}\right|$
for large arguments, the approximation
\begin{equation}
\coth\left(\frac{1}{2}\kappa\left(q_{\mathrm{tot}}\right)t\right)\sim1+2e^{-\kappa\left(q_{\mathrm{tot}}\right)t}\label{eq:coth_approx}
\end{equation}
holds in both, the $t\rightarrow\infty$ and the $q_{\mathrm{tot}}\rightarrow\infty$
limits. Using (\ref{eq:coth_approx}), Eq. (\ref{eq:total_charact_funct})
can be approximated by
\begin{equation}
\tilde{f}\left(q_{\mathrm{tot}}\rightarrow\infty,t\right)\approx\tilde{f}\left(q_{\mathrm{tot}},t\rightarrow\infty\right)\approx\exp\left(-s_{0}\frac{ig_{A}q_{\mathrm{tot}}+\alpha+\kappa\left(q_{\mathrm{tot}}\right)}{\beta}\right).\label{eq:tot_charact_t_inf}
\end{equation}
For $\alpha>0$, this expression approaches $\exp\left(-2\alpha s_{0}/\beta\right)$
as $q_{\mathrm{tot}}\rightarrow0$, whereas for $\alpha\leq0$, it
approaches unity. For the full characteristic function, however, $\tilde{f}\left(q_{\mathrm{tot}},t\right)=1$
is true in all cases. This is depicted in Fig. \ref{fig:charact_funct_approx}.
We conclude that Eq. (\ref{eq:tot_charact_t_inf}) is not a good approximation
for small $\left|q_{\mathrm{tot}}\right|$ if $\alpha>0$ holds. To
find an approximation for small $\left|q_{\mathrm{tot}}\right|$ and
positive growth rates, we expand the numerator and denominator inside
the exponential function in Eq. (\ref{eq:total_charact_funct}) around
$q_{\mathrm{tot}}\approx0$ and find
\[
q_{\mathrm{tot}}s_{0}\left[\alpha+2g_{A}+\kappa\left(q_{\mathrm{tot}}\right)\coth\left(\frac{1}{2}\kappa\left(q_{\mathrm{tot}}\right)t\right)\right]\approx q_{\mathrm{tot}}s_{0}\left[\alpha+2g_{A}+\left|\alpha\right|\coth\left(\frac{\left|\alpha\right|t}{2}\right)\right]
\]
and
\begin{align*}
-i\alpha+g_{A}q_{\mathrm{tot}}-\beta q_{\mathrm{tot}}+i\kappa\left(q_{\mathrm{tot}}\right)\coth\left(\frac{1}{2}\kappa\left(q_{\mathrm{tot}}\right)t\right) & \approx i\left(\left|\alpha\right|\coth\left(\frac{\left|\alpha\right|t}{2}\right)-\alpha\right)+\frac{q_{\mathrm{tot}}\Xi\left(t\right)}{2\left|\alpha\right|},
\end{align*}
where
\begin{align*}
\Xi\left(t\right) & =\text{csch}^{2}\left(\frac{\left|\alpha\right|t}{2}\right)\left[\left|\alpha\right|\left(\beta+g_{A}\left(t\left(\alpha+\beta\right)-1\right)\right)-g_{A}\left(\alpha+\beta\right)\sinh\left(\left|\alpha\right|t\right)+\left|\alpha\right|\left(g_{A}-\beta\right)\cosh\left(\left|\alpha\right|t\right)\right]\\
 & \approx-\beta^{2}\coth\left(\frac{\left|\alpha\right|t}{2}\right)
\end{align*}
In the last line we assumed that $\left|\alpha\right|\ll\beta$. Therefore,
the appropriate approximation for $\alpha>0$ and small $\left|q_{\mathrm{tot}}\right|$
reads
\begin{equation}
\tilde{f}\left(q_{\mathrm{tot}}\rightarrow0,t\right)\approx\exp\left(\frac{q_{\mathrm{tot}}s_{0}\left[\alpha+2g_{A}+\left|\alpha\right|\coth\left(\frac{\left|\alpha\right|t}{2}\right)\right]}{i\left(\left|\alpha\right|\coth\left(\frac{\left|\alpha\right|t}{2}\right)-\alpha\right)+\frac{q_{\mathrm{tot}}\Xi\left(t\right)}{2\left|\alpha\right|}}\right).\label{eq:tot_charact_approx_small_q}
\end{equation}

\begin{figure}
\includegraphics[scale=0.5]{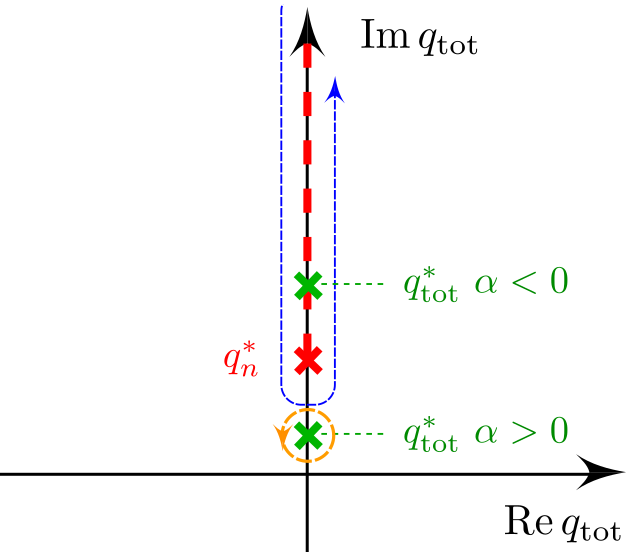}\caption{Branch-cut integration (see Fig. \ref{fig:Branch-cut-integration}
and Eq. (\ref{eq:xs=00003D0_new_contour})) and the pole at $q_{\mathrm{tot}}^{*}$
(Eq. (\ref{eq:q_tot_S-cell_pole})) characterizing the avalanche part
of the distribution. Both, the pole and the branch-cut must be captured
by the deformed contour (blue dashed line). For $\alpha>0$, the pole
at $q_{\mathrm{tot}}^{*}$ is always nearer to the origin than the
starting point of the branch-cut $q_{n}^{*}$ and thus determines
the large $x_{\mathrm{tot}}$ behavior. For $\alpha<0$, $q_{\mathrm{tot}}^{*}$
is within the integration contour, moving towards larger imaginary
values as $t$ is increased. For sufficiently small $t$, the avanlanche
part of the distribution is still well approximated by the integration
around the $q_{\mathrm{tot}}^{*}$ pole (see Fig. \ref{fig:theory_data_powerlaws_avalanches}
b)). \label{fig:Branch-cut_and_pole}}
\end{figure}

It remains to determine the distribution function $f\left(x_{\mathrm{tot}},t\right)$.
At small and intermediate $x_{\mathrm{tot}}$, we expect $f\left(x_{\mathrm{tot}},t\right)$
to be governed by the behavior of the characteristic function at larger
$q_{\mathrm{tot}}$, whereas the behavior of $f\left(x_{\mathrm{tot}},t\right)$
for large $x_{\mathrm{tot}}$ will be determined by the region around
$q_{\mathrm{tot}}\approx0$. This has to do with the analytic structure
of Eq. (\ref{eq:total_charact_funct}). The question is, which nonanalyticity
dominates the Fourier transform as the integration contour is deformed
according to Fig. \ref{fig:Branch-cut-integration}. Besides the branch-cuts
of $\kappa\left(q_{n}\right)$, $\tilde{f}\left(q_{\mathrm{tot}},t\right)$
exhibits a pole on the imaginary axis (see Fig. \ref{fig:Branch-cut_and_pole}).
Let this pole be located at $q_{\mathrm{tot}}=iq_{\mathrm{tot}}^{*}$.
If it is closer to the origin than the upper starting point of the
branch-cut, i.e. $q_{\mathrm{tot}}^{*}<q_{n}^{*}$, where $q_{n}^{*}$
is defined in Eq. (\ref{eq:critical_upper_branch_cut_qn}), the pole
becomes dominant at large $x_{\mathrm{tot}}$. This is because both
nonanalyticities, the branch cut starting at $q_{n}^{*}$ and the
pole at $q_{\mathrm{tot}}^{*}$, are located on the positive imaginary
axis. Hence, when the integration along the deformed contour is performed,
the contributions of the pole and the branch-cut will roughly behave
as $e^{-q_{\mathrm{tot}}^{*}x_{\mathrm{tot}}}$, or $e^{-q_{n}^{*}x_{\mathrm{tot}}}$,
respectively. We can use Eq. (\ref{eq:tot_charact_approx_small_q})
to track the behavior of $q_{\mathrm{tot}}^{*}$ for different parameter
values:
\begin{align}
iq_{\mathrm{tot}}^{*} & =\frac{-2i\left|\alpha\right|}{\Xi\left(t\right)}\left(\left|\alpha\right|\coth\left(\frac{\left|\alpha\right|t}{2}\right)-\alpha\right)\approx2i\frac{\alpha^{2}}{\beta^{2}}\left(1-\mathrm{sign}\left(\alpha\right)\tanh\left(\frac{\left|\alpha\right|t}{2}\right)\right).\label{eq:q_tot_S-cell_pole}
\end{align}
Eq. (\ref{eq:q_tot_S-cell_pole}) shows that the analytic structure
of the characteristic function is very different for positive and
negative $\alpha$. For positive $\alpha$, it is $q_{\mathrm{tot}}^{*}<q_{n}^{*}$,
and therefore $f\left(x_{\mathrm{tot}},t\right)$ will be dominated
by the single pole that is captured in Eq. (\ref{eq:q_tot_S-cell_pole}).
For $\alpha<0$, the opposite scenario is true: we find $q_{n}^{*}<q_{\mathrm{tot}}^{*}$.

We now calculate the inverse Fourier transforms Eqs. (\ref{eq:tot_charact_t_inf})
and (\ref{eq:tot_charact_approx_small_q}) which will give us approximate
expressions for $f\left(x_{\mathrm{tot}},t\right)$ at intermediate
and large $x_{\mathrm{tot}}$. The inverse Fourier transform of Eq.
(\ref{eq:tot_charact_t_inf}) reads
\begin{equation}
f_{\mathrm{tot}}^{t\rightarrow\infty}\left(x_{\mathrm{tot}}\right)\equiv\int\frac{dq_{\mathrm{tot}}}{2\pi}e^{iq_{\mathrm{tot}}x_{\mathrm{tot}}}\lim_{t\rightarrow\infty}\tilde{f}\left(q_{\mathrm{tot}},t\right)=e^{-\frac{\alpha}{\beta}s_{0}}\int\frac{dq_{n}}{2\pi}\exp\left(iq_{\mathrm{tot}}\left(x_{\mathrm{tot}}-s_{0}\frac{g_{A}}{\beta}\right)-s_{0}\frac{\kappa\left(q_{\mathrm{tot}}\right)}{\beta}\right).\label{eq:fourier_total_charact_t_inf}
\end{equation}
We encountered this integral when we calculated the distribution of
N-cells in Eq. (\ref{eq:fully_evolved_n_distr}). Similarly to Eq.
(\ref{eq:evolved_lineages_powerlaw}) we find
\begin{equation}
f_{\mathrm{tot}}^{t\rightarrow\infty}\left(x_{\mathrm{tot}}\right)=\frac{s_{0}\sqrt{p}e^{\frac{-ps_{0}^{2}}{2\beta^{2}\left(x_{\mathrm{tot}}-s_{0}\frac{g_{A}}{\beta}\right)}-q_{\mathrm{tot}}^{*}\left(x_{\mathrm{tot}}-s_{0}\frac{g_{A}}{\beta}\right)}}{\sqrt{2\pi}\beta\left(x_{\mathrm{tot}}-s_{0}\frac{g_{A}}{\beta}\right)^{3/2}}\label{eq:f_tot_x_tot_full_N=00003D1}
\end{equation}
with 
\begin{align*}
p & =\sqrt{\beta g_{A}\left(2\alpha g_{A}+\alpha^{2}+\beta g_{A}\right)}\\
q_{n}^{*} & =\frac{\alpha^{2}}{g_{A}\left(\alpha+\beta\right)+\sqrt{\beta g_{A}\left(\alpha^{2}+2\alpha g_{A}+\beta g_{A}\right)}}.
\end{align*}
In order to account for the $\bar{N}$ N-cells that each S-cell is
producing we have to make the substitution $q_{\mathrm{n}}\rightarrow\bar{N}q_{n}$
in Eq. (\ref{eq:charact_q->qtot}). This substitution carries over
to Eq. (\ref{eq:tot_charact_t_inf}) where we have to replace $q_{\mathrm{tot}}$
by $\bar{N}q_{\mathrm{tot}}$. The integration of Eq. (\ref{eq:fourier_total_charact_t_inf})
is evaluated according to
\begin{equation}
\int\frac{dq_{\mathrm{tot}}}{2\pi}e^{iq_{\mathrm{tot}}x_{\mathrm{tot}}}\lim_{t\rightarrow\infty}\tilde{f}\left(\bar{N}q_{\mathrm{tot}},t\right)=\int\frac{dq'_{\mathrm{tot}}}{\bar{N}2\pi}e^{iq'_{\mathrm{tot}}x_{\mathrm{tot}}/\bar{N}}\lim_{t\rightarrow\infty}\tilde{f}\left(q'_{\mathrm{tot}},t\right)=\frac{1}{\bar{N}}f_{\mathrm{tot}}^{t\rightarrow\infty}\left(x_{\mathrm{tot}}/\bar{N}\right).\label{eq:substitute_N_after_int}
\end{equation}
 For $\left|\alpha\right|\ll\beta$, we obtain the result of Eq. (\ref{eq:critical_Levy_distr}),
Sec. \ref{sec:Tissue-growth:-power-laws} of the main text:
\begin{equation}
f_{\mathrm{tot}}^{t\rightarrow\infty}\left(x_{\mathrm{tot}}\right)\approx\frac{s_{0}\beta e^{-\frac{\alpha^{2}x_{\mathrm{tot}}}{\beta^{2}\bar{N}}}}{2\sqrt{2\pi}\bar{N}\left(x_{\mathrm{tot}}/\bar{N}\right)^{3/2}}.\label{eq:append_final_power_law}
\end{equation}
According to the above discussion, Eq. (\ref{eq:append_final_power_law})
is valid for intermediate $x_{\mathrm{tot}}$ if $\alpha>0$. For$\alpha<0$,
the contribution of the pole becomes stronger and stronger sub-leading
to the branch-cut contribution as $t$ is increased (see Eq. (\ref{eq:q_tot_S-cell_pole})).
However, if $t$ is sufficiently small, the pole is in the vicinity
of the branch-cut starting point $q_{n}^{*}$ and has a strong influence
on the distribution function. For large enough $t$ however, Eq. (\ref{eq:append_final_power_law})
is a good approximation for $f_{\mathrm{tot}}\left(x_{\mathrm{tot}}\right)$
everywhere.

To determine the behavior at large $x_{\mathrm{tot}}$ for $\alpha>0$
we make use of Eq. (\ref{eq:tot_charact_approx_small_q}) which gives
a good approximation for the characteristic function at small $q_{\mathrm{tot}}$.
It is useful to rewrite Eq. (\ref{eq:append_final_power_law}) as
\[
\tilde{f}\left(q_{\mathrm{tot}}\rightarrow0,t\right)=e^{-aq_{\mathrm{tot}}^{*}}\exp\left(\frac{aq_{\mathrm{tot}}^{*}}{1+iq_{\mathrm{tot}}/q_{\mathrm{tot}}^{*}}\right)
\]
with
\begin{equation}
a=\frac{s_{0}\left(\alpha+2g_{A}+\alpha\coth\left(\frac{\alpha t}{2}\right)\right)}{\alpha\left(\coth\left(\frac{\alpha t}{2}\right)-1\right)}.\label{eq:a_def}
\end{equation}
The inverse Fourier transform
\begin{equation}
f_{\mathrm{tot}}\left(x_{\mathrm{tot}}\rightarrow\infty,t\right)=e^{-aq_{\mathrm{tot}}^{*}}\int\frac{dq_{\mathrm{tot}}}{2\pi}\exp\left(iq_{\mathrm{tot}}x_{\mathrm{tot}}+\frac{aq_{\mathrm{tot}}^{*}}{1+iq_{\mathrm{tot}}/q_{\mathrm{tot}}^{*}}\right)\label{eq:integral_around_pole}
\end{equation}
is strictly speaking divergent (as is the one in Eq. (\ref{eq:charact_funct_delta_manipulation})).
However it can be regularized by adding a small exponential factor
$e^{-q_{\mathrm{tot}}\left|\varepsilon\right|}$ to the integrand
and letting $\varepsilon\rightarrow0$ after the integral is done.
The large $x_{\mathrm{tot}}$ behavior that we are interested in is
dominated by the pole at $q_{\mathrm{tot}}^{*}$ for $\alpha>0$.
We will see later, that for $\alpha<0$, the pole still dominates
the distribution for reasonably large $x_{\mathrm{tot}}$ (avalanche
regime) if $t$ is small, although it does not govern the asymptotics
for $x_{\mathrm{tot}}\rightarrow\infty$ . The pole's contribution
can be found by integrating along a circle around $q_{\mathrm{tot}}^{*}$
(see Fig. \ref{fig:Branch-cut_and_pole}). Writing
\[
q_{\mathrm{tot}}=iq_{\mathrm{tot}}+re^{i\varphi}
\]
the integral around the pole at $q_{\mathrm{tot}}$ becomes
\begin{align*}
f_{\mathrm{tot}}\left(x_{\mathrm{tot}}\rightarrow\infty,t\right) & =\exp\left(-\frac{a}{q_{\mathrm{tot}}}-\frac{x_{\mathrm{tot}}}{q_{\mathrm{tot}}}\right)ir\int_{0}^{2\pi}\frac{e^{i\varphi}d\varphi}{2\pi}\exp\left(ire^{i\varphi}x_{\mathrm{tot}}+\frac{a}{iq_{\mathrm{tot}}^{2}re^{i\varphi}}\right).
\end{align*}
Since the radius $r$ is arbitrary, we are free to chose
\[
r=\sqrt{\frac{a}{q_{\mathrm{tot}}^{2}x_{\mathrm{tot}}}}.
\]
After some algebra, we arrive at 
\begin{equation}
f_{\mathrm{tot}}\left(x_{\mathrm{tot}}\rightarrow\infty,t\right)=\exp\left(-\frac{a}{q_{\mathrm{tot}}}-\frac{x_{\mathrm{tot}}}{q_{\mathrm{tot}}}\right)r\int_{0}^{2\pi}\frac{e^{i\varphi}d\varphi}{2\pi}\exp\left(rx_{\mathrm{tot}}\left[e^{i\varphi}+\frac{1}{e^{i\varphi}}\right]\right).\label{eq:contour_0_2pi}
\end{equation}
Using the integral representation of Bessel functions
\[
I_{n}\left(z\right)=\frac{1}{2\pi i}\int_{\mathcal{C}}dt\frac{e^{\frac{z}{2}\left(t+\frac{1}{t}\right)}}{t^{n+1}},
\]
where $\mathcal{C}$ is a contour enclosing the origin, with the substitution
$t=e^{i\varphi}$ Eq. (\ref{eq:contour_0_2pi}) becomes
\begin{equation}
f_{\mathrm{tot}}\left(x_{\mathrm{tot}}\rightarrow\infty,t\right)=\theta\left(x_{s}\right)\exp\left(-\frac{a}{q_{\mathrm{tot}}}-\frac{x_{\mathrm{tot}}}{q_{\mathrm{tot}}}\right)\sqrt{\frac{a}{q_{\mathrm{tot}}^{2}x_{\mathrm{tot}}}}I_{1}\left(2\sqrt{\frac{ax_{\mathrm{tot}}}{q_{\mathrm{tot}}^{2}}}\right).\label{eq:Bessel_approximation_to_avalanche}
\end{equation}
For $\alpha>0$, this is the large $x_{\mathrm{tot}}$ approximation
for the distribution function $f\left(x_{\mathrm{tot}},t\right)$
given in Eq. (\ref{eq:avalanche_approx}), where we have restored
the variable $\bar{N}$. As is demonstrated in Fig. (\ref{fig:theory_data_powerlaws_avalanches}),
Eq. (\ref{eq:Bessel_approximation_to_avalanche}) provides a good
approximation for the large $x_{\mathrm{tot}}$ behavior of $f_{\mathrm{tot}}\left(x_{\mathrm{tot}},t\right)$
for a positive S-cell growth rate $\alpha$. However, even for $\alpha<0$,
at reasonably large $x_{\mathrm{tot}}$ (namely in the avalanche regime),
the behavior of the lineage size probability density is well described
by Eq. (\ref{eq:Bessel_approximation_to_avalanche}), if $\alpha t\lesssim1$
holds. This is due to the fact that the pole at $q_{\mathrm{tot}}^{*}$,
while located above $q_{n}^{*}$ on the imaginary $q_{\mathrm{tot}}$
line (see Fig. \ref{fig:Branch-cut_and_pole}), is still sufficiently
near $q_{n}^{*}$to dominate the avalanche part of the distribution
function (see Eq. (\ref{eq:q_tot_S-cell_pole})). However, the asymptotics
of $f_{\mathrm{tot}}\left(x_{\mathrm{tot}},t\right)$ for $x_{\mathrm{tot}}\rightarrow\infty$
will not be given by Eq. (\ref{eq:Bessel_approximation_to_avalanche}).
Since in both cases, for positive and negative $\alpha$, Eq. (\ref{eq:Bessel_approximation_to_avalanche})
is an approximation for the avalanche part of the lineage size probability
density, we call it $f_{\mathrm{tot}}^{\mathrm{av.}}\left(x_{\mathrm{tot}},t\right)$.

\subsection{Most-likely paths\label{sec:Most-likely-paths}}

We now derive the most likely path $s_{\tau}(t)$ that a critical
S-cell population (i.e. $\alpha=0$) takes from $s_{0}$ cells at
$t=0$ to extinction at $t=\tau$. Since we consider only S-cells
and only critical populations, equation (\ref{eq:Fokker-Planck}),
accorting to the rules of Ito's calculus, reduces a process described
by the stochastic differential equation (SDE)
\begin{equation}
ds=\sqrt{{\beta s}}dw(t)\label{eq:just_S_critical_SDE}
\end{equation}
where $w(t)$ is the standard Brownian motion. If the set of possible
paths was finite-dimensional, we could proceed by finding the density
$W$ defined by equation (\ref{eq:just_S_critical_SDE}) on the set
of possible paths starting at $s_{0}$ and maximizing $W(s_{\tau})$
subject to $s_{\tau}(\tau)=0$ to find $s_{\tau}$. It turns out,
however, that this approach only cleanly generalizes to infinite-dimensional
path spaces in the case of a constant diffusion term \citep{durr_onsager-machlup_1978}.
To avoid these technical difficulties, we perform a change of variable
to transform equation (\ref{eq:just_S_critical_SDE}) into a process
with a constant diffusion term. By setting $y(t)=2\sqrt{s\left(t\right)/\beta}$
and applying Itô's lemma we get 
\begin{equation}
dy=A(y)dt+dw(t)\text{{\ where\ }A(y)=\ensuremath{-\frac{{1}}{2y}}}\label{eq:y_SDE}
\end{equation}
(were once $y$ reaches zero, we define it to remain there, despite
$A$ becoming undefined). For this process, the density functional
$W$ expressed in terms of its Lagrangian $L$ (also called Onsager-Machlup
function) is \citep{durr_onsager-machlup_1978}
\begin{eqnarray}
W[y(t)] & \sim & e^{-S(y)}\label{eq:y_W}\\
\text{{where\ }}S\left[y\right] & = & \frac{{1}}{2}\int_{0}^{T}L(y,\dot{y})dt\nonumber \\
\text{{and\ }}L(y,\dot{y}) & = & (\dot{y}-A(y))^{2}+A'(y).\nonumber 
\end{eqnarray}

The functional $W$ defines a probability density on the set of paths
in the following sense: The probability for a random path to lie within
a small tube of diameter $\epsilon$ around a given differentiable
path $y$ is asymptotically proportional to $W[y]H(\epsilon)$ for
some function $H$ independent of $y$ (this factorization fails if
the diffusion term is not constant \citep{durr_onsager-machlup_1978}).
We can therefore proceed as we would in the finite-dimensional case
and maximize $W$ to find $y_{\tau}$. Since maximizing $W$ means
minimizing the \emph{action} $S(y)=\frac{{1}}{2}\int_{0}^{T}L(y,\dot{y)dt}$,
the desired $y_{\tau}$ is found by solving the Euler-Lagrange (EL)
equation $\frac{{d}}{dt}\frac{{\partial L}}{\partial\dot{y}}=\frac{{\partial L}}{\partial y}$.
In our case the EL equation yields $\ddot{y}=-\frac{{3}}{4}y^{-3}$
with the general solution
\begin{eqnarray}
y(t) & = & \sqrt{{\lambda t^{2}+\text{\ensuremath{\mu t}+\ensuremath{\nu}}}}\label{eq:y_solution}\\
\textrm{{where\ }}\mu^{2} & = & 4\lambda\nu+3,\nonumber 
\end{eqnarray}
and solving for the boundary conditions $y(0)=u(s_{0})=2\sqrt{{s_{0}/\beta}}$
and $y(\tau)=0$ yields
\begin{alignat*}{2}
\nu= & s_{0}\frac{{4}}{\beta}, & \mu=\sqrt{{3}}-2\frac{{\nu}}{\tau}, & \lambda=\frac{{1}}{\tau}\left(\frac{{\nu}}{\tau}-\sqrt{3}\right)
\end{alignat*}
(where we chose the solution that ensures $\lambda t^{2}+\text{\ensuremath{\mu t}+\ensuremath{\nu}}>=0$
on $[0,\tau]$). Inserting these parameters into the general solution
(\ref{eq:y_solution}) and transforming back from $y$ to $s$ produces
after some rearrangements the extinction trajectory stated in equation
(\ref{eq:extinct_traj}).

\subsection{Full Fokker-Planck equation and the simplification of the A$\rightarrow$A+N
process}

The full Fokker-Planck equation for the SAN process of Fig \ref{fig:SAN_model}
of the main text is given by
\begin{eqnarray}
\frac{\partial f\left(\mathbf{x},t\right)}{\partial t} & = & -\alpha\frac{\partial}{\partial s}sf\left(\mathbf{x},t\right)+\frac{\beta}{2}\frac{\partial^{2}}{\partial s^{2}}sf\left(\mathbf{x},t\right)\nonumber \\
 &  & -\frac{\partial}{\partial a}g_{A}sf\left(\mathbf{x},t\right)+\frac{1}{2}\frac{\partial^{2}}{\partial a^{2}}g_{A}sf\left(\mathbf{x},t\right)-g_{A}\frac{\partial}{\partial s}\frac{\partial}{\partial a}x_{s}f\left(\mathbf{x},t\right).\nonumber \\
 &  & -g_{N}\frac{\partial}{\partial n}af\left(\mathbf{x},t\right)+\frac{g_{N}}{2}\frac{\partial^{2}}{\partial n^{2}}af\left(\mathbf{x},t\right),\label{eq:Full_Fokker-Planck}
\end{eqnarray}
where $\mathbf{x}=\left(s,a,n\right)$ is the vector of S, A, and
N-cell counts. The simplification of the A$\rightarrow$A+N process
used in our calculations is that each A-cell, on average, produces
$\bar{N}$ N-cells. This, essentially, amounts to cutting the last
line of the above equation and rescaling the the $a$ variable by
a factor of $\bar{N}$, giving Eq. (\ref{eq:Fokker_Planck_operator})
of the main text.\lyxdeleted{Egor}{Tue Jun 20 11:40:34 2023}{ }

Our justification for this simplification is that, while fluctuations
in the number of S-cells $s$ scale as $\sim\sqrt{s}$ (see also Eq.
(\ref{eq:just_S_critical_SDE})), the fluctuations of the N-cell count
are - for a given $a$ - Gaussian. This can be estimated from the
diffusion term in the last line. While the large $\sim\sqrt{s}$ fluctuations
of the stem-cell number ultimately lead to the power-law at criticality,
the Gaussian fluctuations of the number of neurons lead only to a
broadening of the distributions. Llorca et. al. - eLife, 2019 (see
main text for the full reference) considered the neuronal output of
precurser cells (radial glia cells, which correspond to our A cells),
and found that its uncertainty is only about an order of magnitude
in size. As this is uncertainty is small compared to the uncertainty
of the power-law distribution of total cell counts, we choose the
simplified model, which is analytically tractable.

\subsection{Kolmogorov-Smirnov tests of the lineage size distribution\label{subsec:Kolmogorov-Smirnov-tests}}

To ascertain our results, we performed Kolmogorov-Smirnov tests of
the empirical data and our theoretical results for the pdf at day
40. Since our conclusion that the growth process is close to criticality
is true for any parameter values within the given error margins of
our fits (Table \ref{tab:parameter_table}), but the K-S statistic
is expected to strongly depend on which values are chosen ( as is
obvious from a visual evaluation of the error margins of Fig. 2),
we believe that what should be tested, is the distribution for intermediate
lineage sizes (say between $\sim200$ and $\sim2000$). Here, the
distribution is both close to a 3/2 power-law and mostly independent
on the dynamical parameters of the model. As expected, the precise
p-values of the K-S test strongly depend on the interval under consideration,
but they support, in general, the SAN-model hypothesis ($p>0.1$).
The highest p-values are obtained for lineage sizes between 450 and
1500. Here, the p-value is $p=0.81$, with the data comprising 1402
lineages.

The following plot shows the p-values of all intervals between $x_{\mathrm{tot}}=100$
and $x_{\mathrm{tot}}=4000$. As expected, we find a cluster of significant
p values between $x_{\mathrm{tot}}\approx200$ and $x_{\mathrm{tot}}\approx2000$,
corresponding to the region where the power-law is most pronounced.
Surprisingly, there is also a cluster of high p-values between $x_{\mathrm{tot}}\approx2000$
and $x_{\mathrm{tot}}\approx3500$ (this region is comprised of $\sim400$
data points) -- corresponding to the region containing the avalanche
of active s-cells and the exponential truncation at large $x_{\mathrm{tot}}$.
In contrast, the data in the crossover between the power-law region
and the truncation region does not produce high p values. We assume,
that the uncertainties and systematic errors of our model have more
weight in this region of the distribution.

\begin{figure}
\centering{}\includegraphics[scale=0.6]{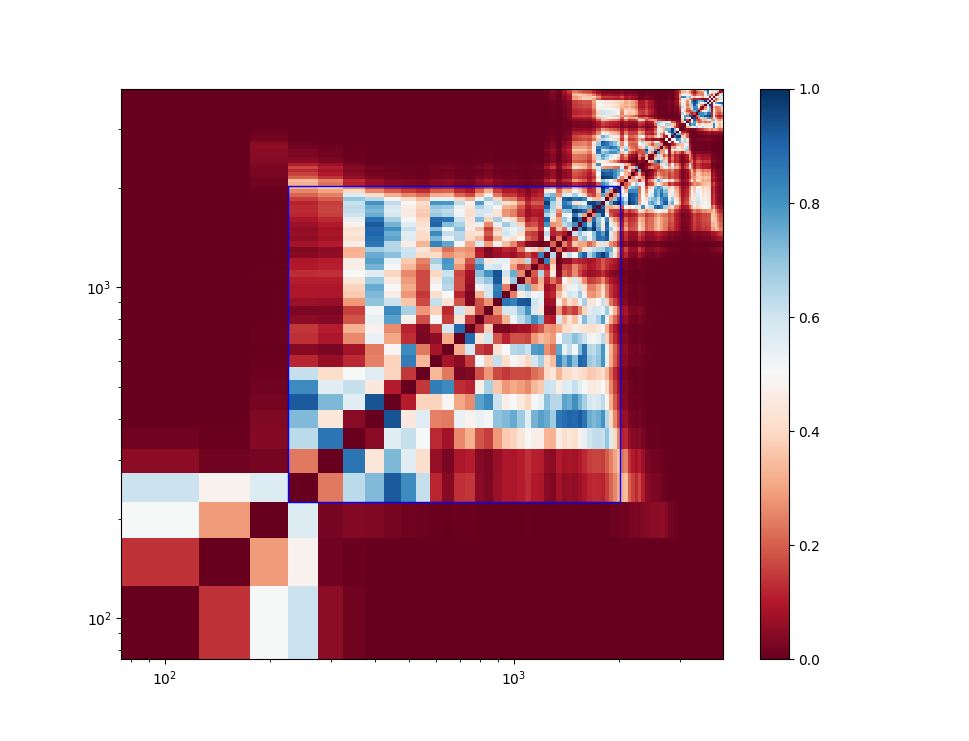}\caption{$p$-values of the Kolmogorov-Smirnov test comparing the theoretical
probability distribution of the SAN model to the data at day 40 on
different intervals along the axis of total cell counts $x_{\mathrm{tot}}$.
The figure should be read as follows: choose an interval between $x_{\mathrm{tot,min}}$
and $x_{\mathrm{tot,max}}$ in which the data should be compared to
the theoretical distribution. The color map at the point $x=x_{\mathrm{tot,min}}$
and $y=x_{\mathrm{tot,max}}$ (or the other way around) gives the
$p$-value on the given interval. Intervals between 225 and 2000 -
the cell counts for which the power-law is most pronounced - lie within
the blue rectangle.}
\end{figure}

\subsection{Histograms of the data for different bin sizes}

In the following the show a comparison between the probability density
functions of the SAN model and histograms of the total cell count
data for different bin sizes. We find that the agreement is largely
independent of the bin size. The plots correspond to Fig \ref{fig:theory_data_powerlaws_avalanches}
a) of the main text, with altered bin sizes for the data.\lyxdeleted{Egor}{Tue Jun 20 11:40:34 2023}{ }

\begin{figure}[H]
\centering{}\includegraphics[scale=0.35]{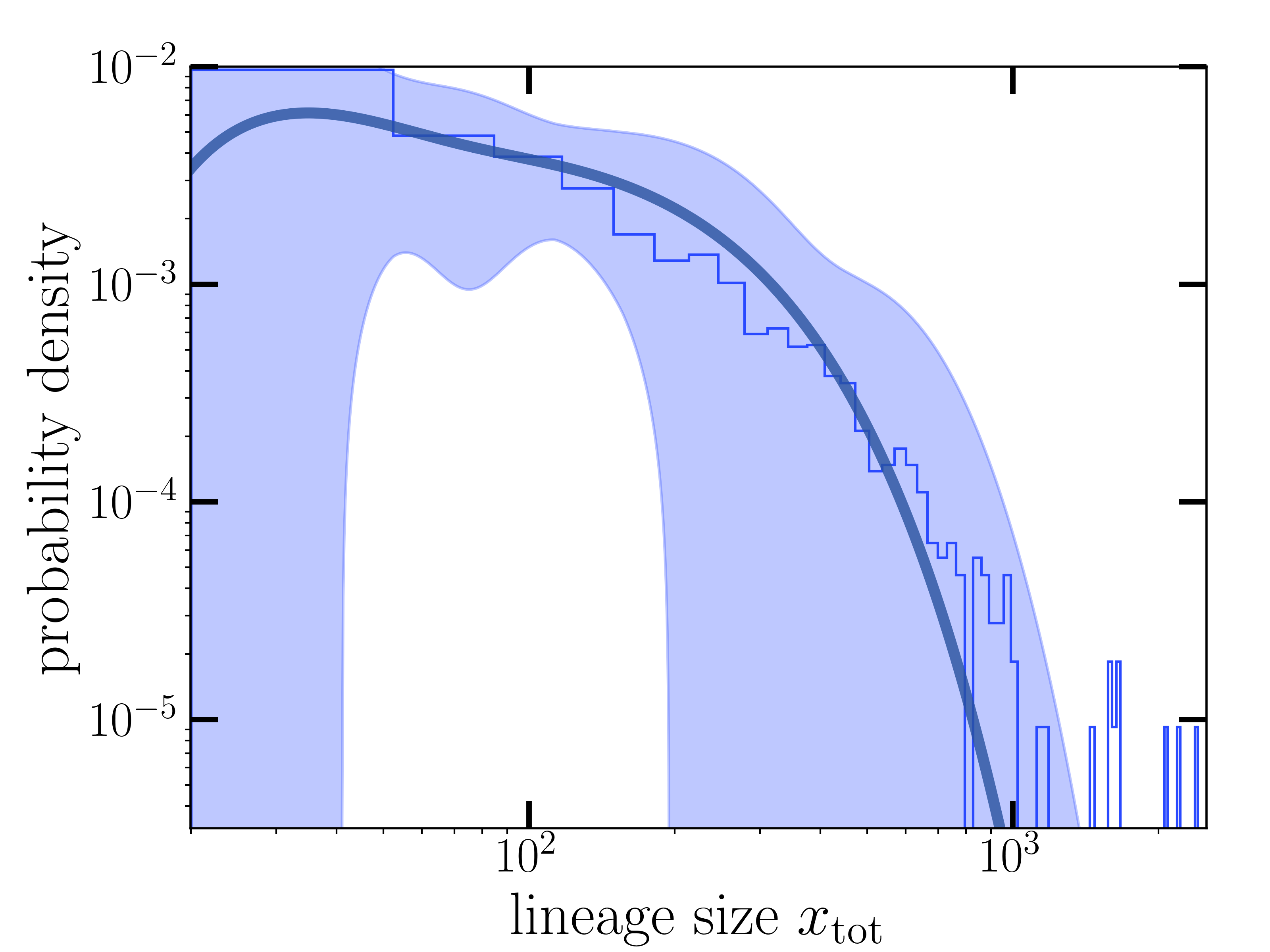}$\hspace{1em}$\includegraphics[scale=0.35]{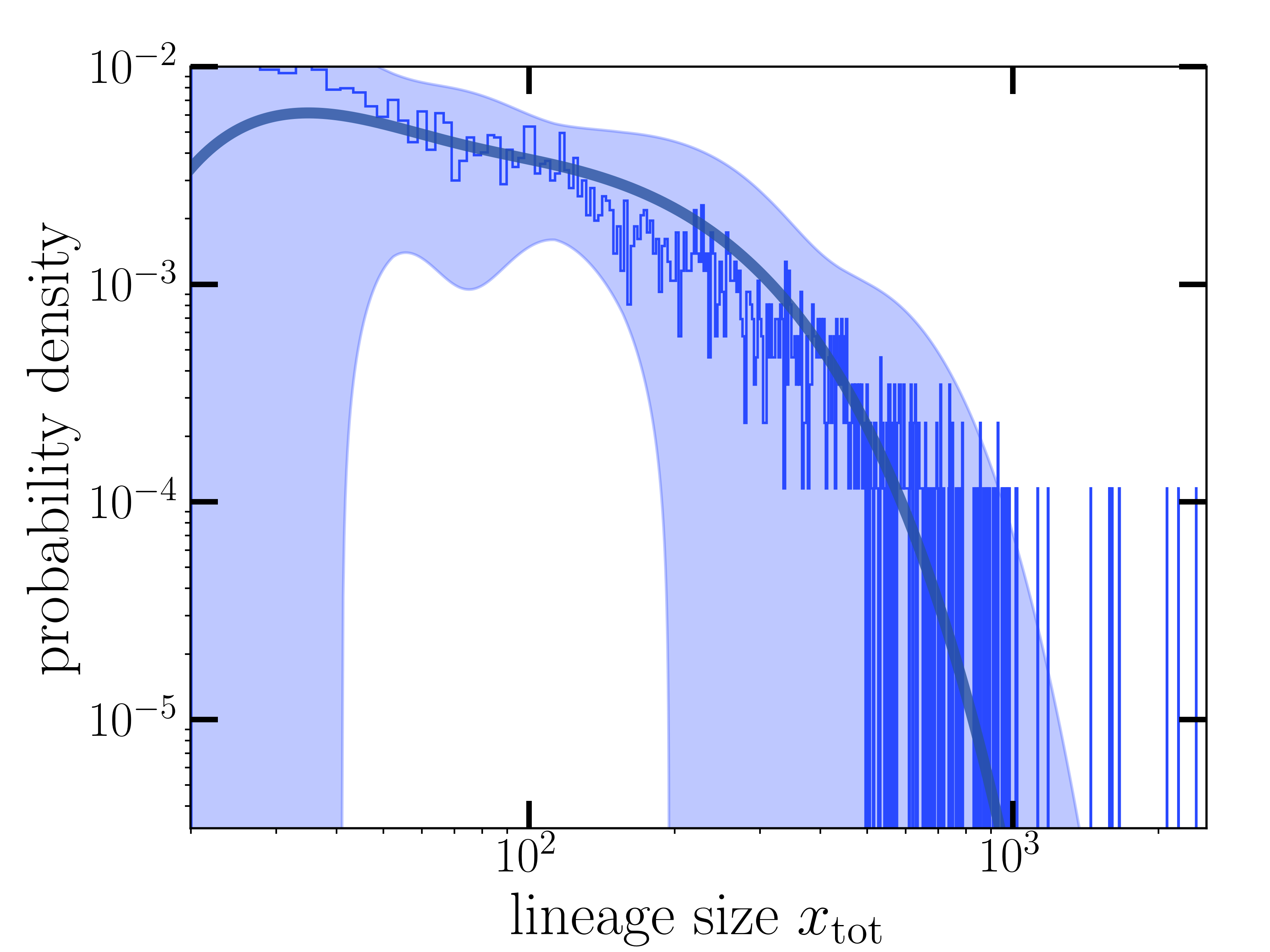}$\hspace{1em}$\includegraphics[scale=0.35]{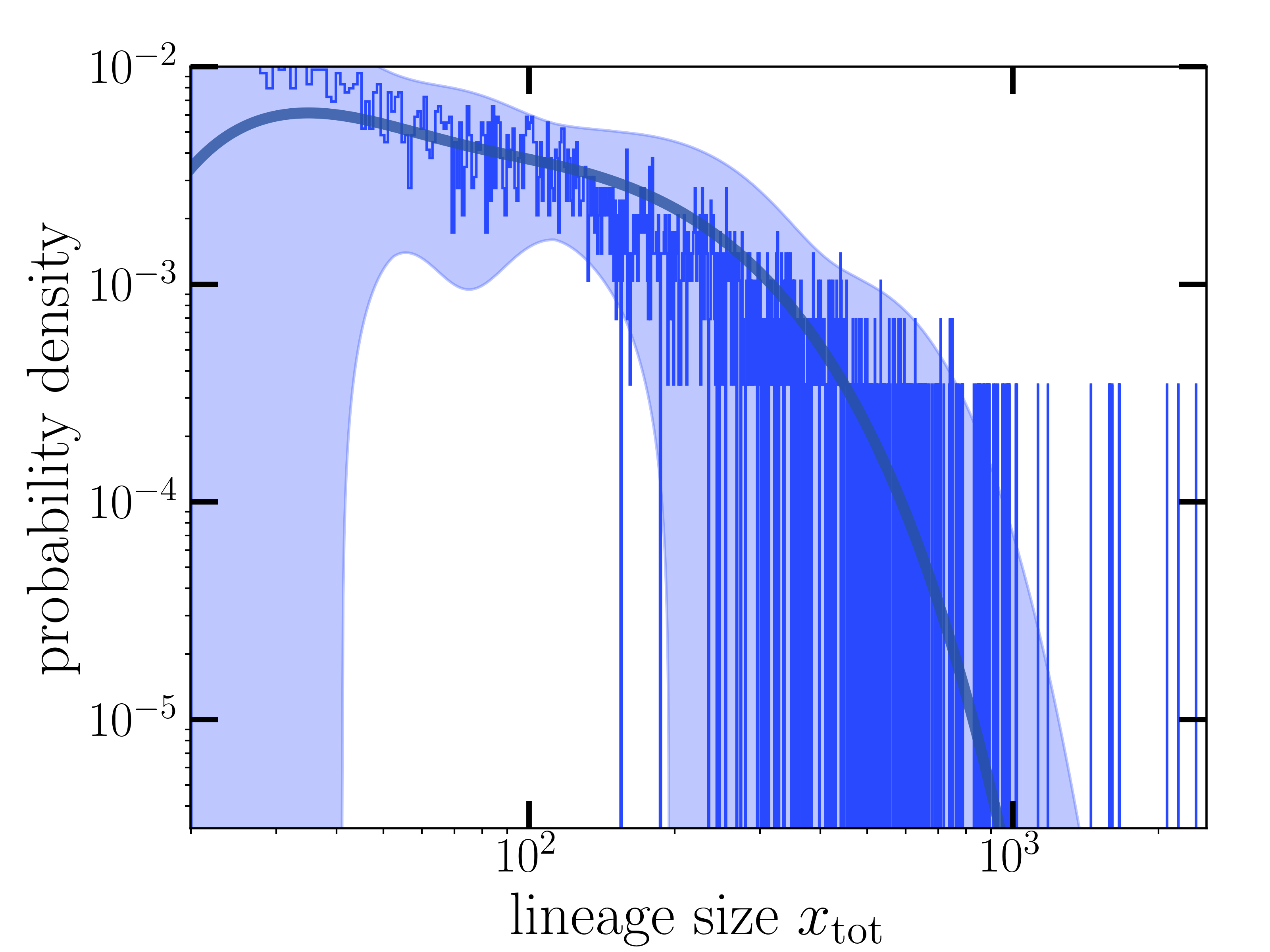}\caption{Histograms of the data for an organoid at day 16 with bin sizes 80,
1000 and 3000 compared to the SAN-model predictions for the total
lineage size probability densites. Shaded areas show the uncertainties
of the fitted parameters (see Table \ref{tab:parameter_table} of
the main text).}
\end{figure}

\begin{figure}[H]
\centering{}\includegraphics[scale=0.35]{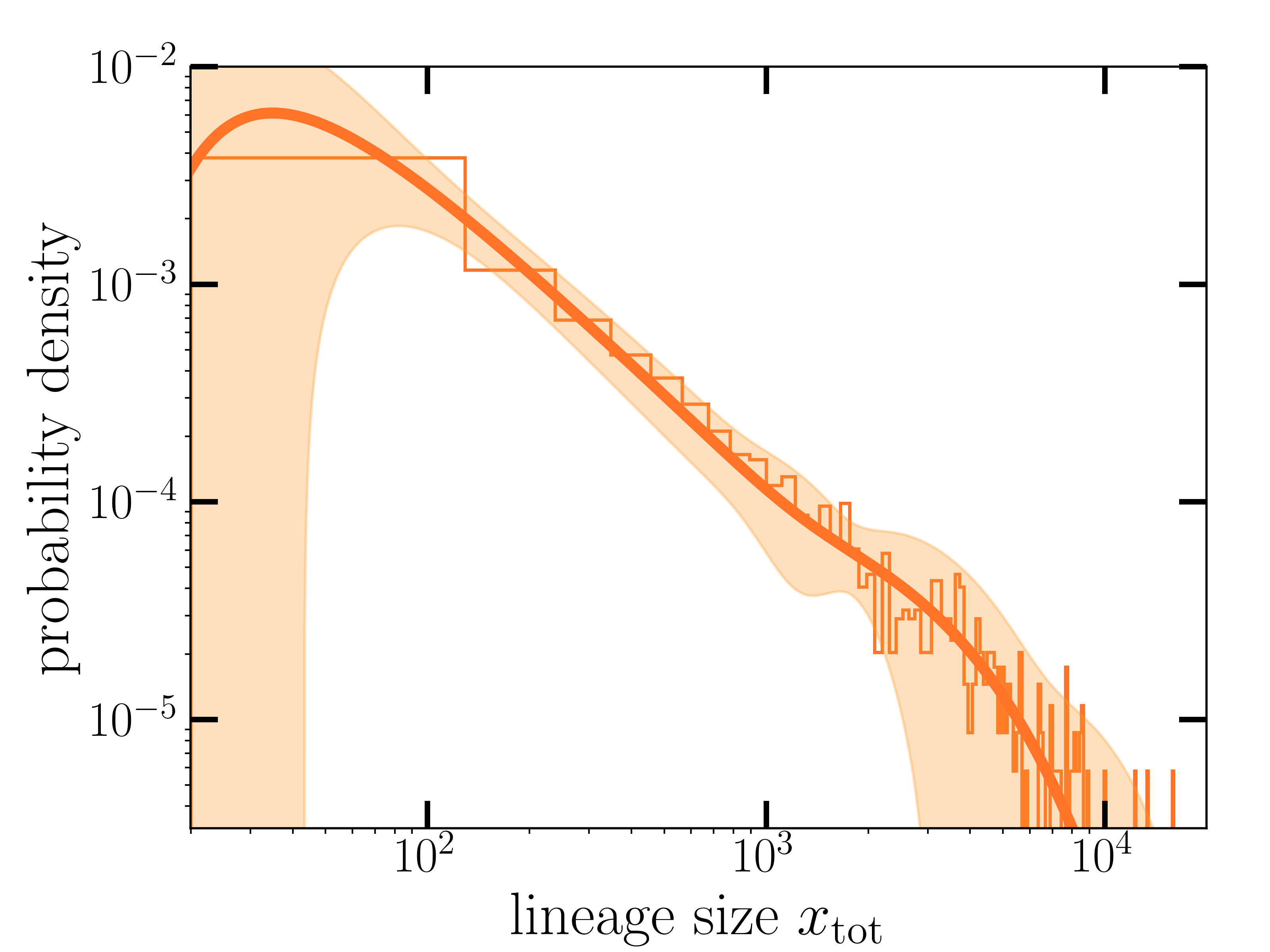}$\hspace{1em}$\includegraphics[scale=0.35]{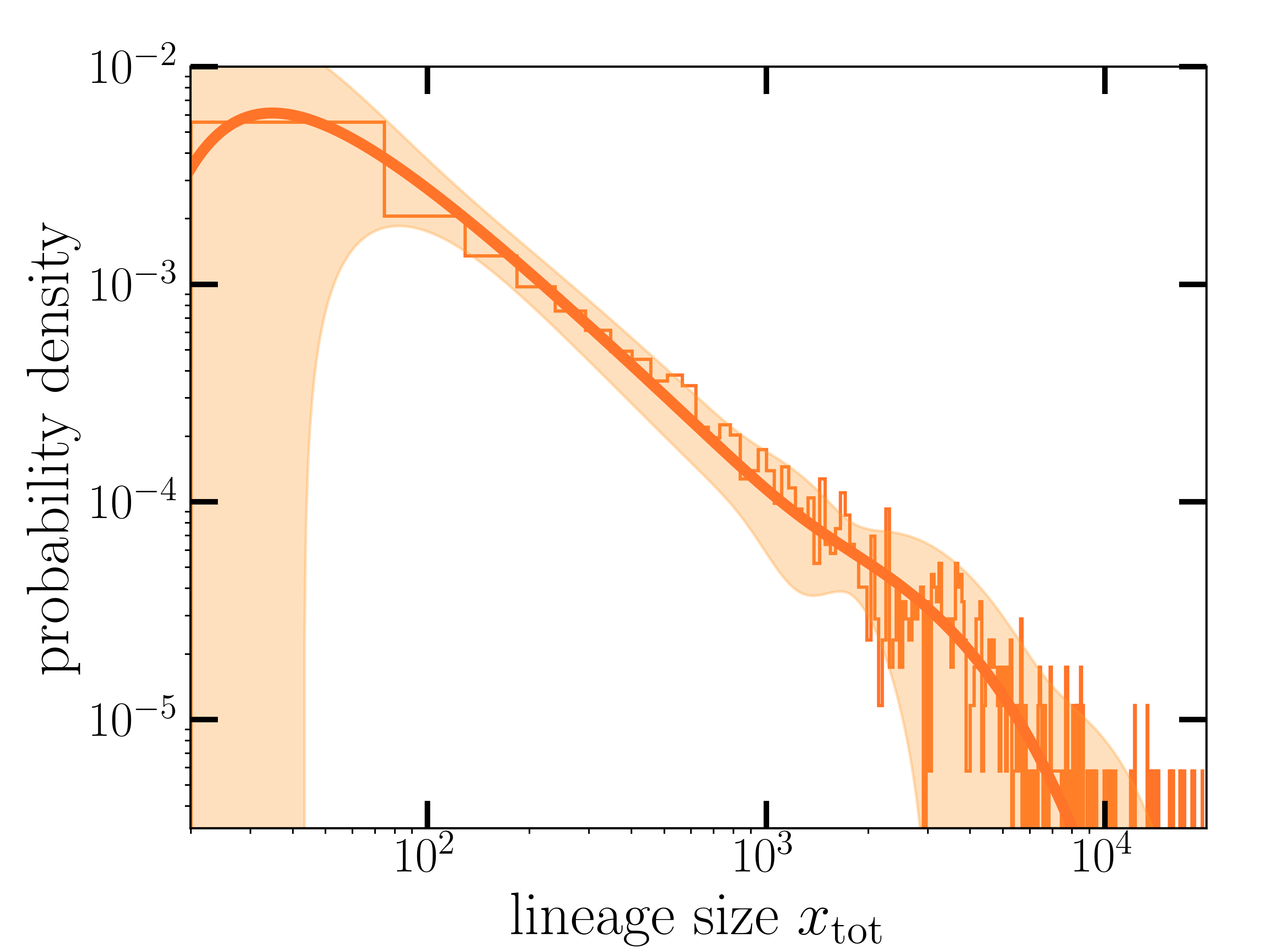}$\hspace{1em}$\includegraphics[scale=0.35]{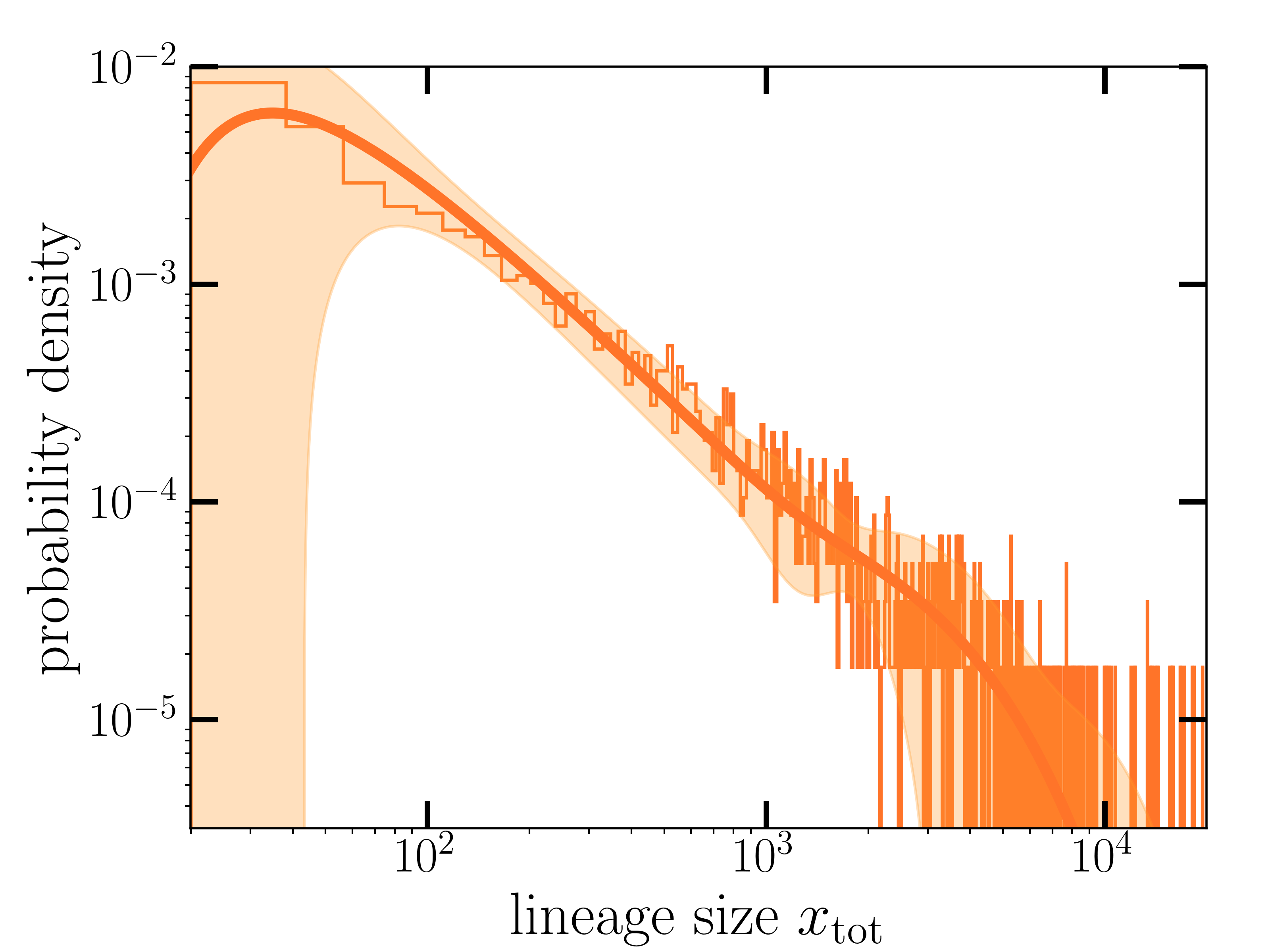}\caption{Histograms of the data for an organoid at day 40 with bin sizes 500,
1000 and 3000 compared to the SAN-model predictions for the total
lineage size probability densites. Shaded areas show the uncertainties
of the fitted parameters (see Table \ref{tab:parameter_table} of
the main text).}
\end{figure}

\end{document}